\documentclass[aps,prd,superscriptaddress,twoside,twocolumn,nofootinbib,showpacs,floatfix]{revtex4-1}
\usepackage{amsmath,amssymb}
\usepackage{graphicx,bm,braket}
\usepackage{subfigure}
\pdfpagewidth=\paperwidth
\pdfpageheight=\paperheight
\allowdisplaybreaks
\usepackage{epstopdf}
\newcommand*{\slashed}[1]{{#1\!\!\!/}}

\begin{document}

\title{\boldmath Nucleon resonances in $\gamma p \to \omega p$ reaction}

\author{N. C. Wei}
\affiliation{Department of Physics, Zhengzhou University, Zhengzhou, Henan 450001, China}
\affiliation{School of Nuclear Science and Technology, University of Chinese Academy of Sciences, Beijing 100049, China}

\author{F. Huang}
\email{huangfei@ucas.ac.cn}
\affiliation{School of Nuclear Science and Technology, University of Chinese Academy of Sciences, Beijing 100049, China}

\author{K. Nakayama}
\email{nakayama@uga.edu}
\affiliation{Department of Physics and Astronomy, University of Georgia, Athens, Georgia 30602, USA}

\author{D. M. Li}
\email{lidm@zzu.edu.cn}
\affiliation{Department of Physics, Zhengzhou University, Zhengzhou, Henan 450001, China}

\date{\today}

\begin{abstract}
The most recent high-precision data on spin observables $\Sigma$, $T$, $P'$, $E$, $F$ and $H$ reported by the CLAS Collaboration together with the previous data on differential cross sections and spin-density-matrix elements reported by the CLAS, A2, GRAAL, SAPHIR and CBELSA/TAPS Collaborations for the reaction $\gamma p \to \omega p$ are analyzed within an effective Lagrangian approach. The reaction amplitude is constructed by considering the $t$-channel $\pi$ and $\eta$ exchanges, the $s$-channel nucleon and nucleon resonances exchanges, the $u$-channel nucleon exchange and the generalized contact current. The latter accounts effectively for the interaction current and ensures that the full photoproduction amplitude is gauge invariant. It is shown that all the available CLAS data can be satisfactorily described by considering the $N(1520)3/2^-$, $N(1700)3/2^-$, $N(1720)3/2^+$, $N(1860)5/2^+$, $N(1875)3/2^-$, $N(1895)1/2^-$ and $N(2060)5/2^-$ resonances in the $s$-channel. The parameters of these resonances are extracted and compared with those quoted by PDG.
\end{abstract}



\maketitle

\section{Introduction} \label{introduction}

The study of hadron mass spectrum and hadronic decays provides essential information towards the understanding of strong interaction in the non-perturbative regime of Quantum Chromodynamics. It is now a consensus in hadron physics community that one has to investigate as many independent reaction processes as possible to extract detailed information on hadron resonances, especially, in the baryonic sector. Indeed, there is currently an intense activity, both experimentally and theoretically, in investigating many different meson production reactions. One of the major motivations behind this drive is to find the so-called ``missing resonances", which are predicted by the non-relativistic quark models but not found in the experiments of $\pi$ production reactions \cite{Isgur:1977ef,Koniuk:1979vy}. A possible explanation for the ``missing resonances" problem is that they may have escaped from observation due to their relatively small coupling to the $\pi N$ final state. Thus the study of production reactions of mesons other than $\pi$ becomes indispensable. The $\eta N$ channel has been investigated as a first step towards this goal and, currently, heavier meson production processes such as $\eta'$, $\omega$ and $\phi$ are being subjects of increasing attention. These efforts are not restricted to the non-strangeness sector only. There has been also an intense activity in the strangeness sector to search for hyperon resonances with strangeness quantum number $S=-1$ via $KY$ photoproduction ($Y=\Lambda$, $\Sigma$) and interest in heavier meson photoproduction such as $K^*Y$ is also increasing \cite{Wang:2017tpe,Wang2018,Wang2019,Wei2019,Kim:2011,Kim:2014}. There are also initiatives to investigate strange baryon resonances with $S=-2$ and $-3$ \cite{Jackson:2015,Ryu:2013}.

The present work concerns the photoproduction of $\omega$ meson off the proton. This reaction has been studied intensively from late 60ths to 90ths at energies well above the resonance energy region to address a variety of interesting physics questions. In particular, the vector meson photo- and electro-production processes, in general,  provide an important insight into the diffractive mechanisms at high energies. At lower energies in the resonance energy region (below center-of-mass energy of $W \sim 3$ GeV), the $\omega$ photoproduction offers a means of probing the possible missing states in the 2 GeV mass region (referred to as the third resonance region) which might couple to the $\omega N$ channel. The $\omega$ photoproduction off nuclei has been also investigated \cite{Trnka:2005,Kaskulov:2007} to study the medium effects on the properties of vector mesons \cite{Brown:1991,Hatsuda:1992}. The $\omega$ photoproduction off deuteron has been investigated in Ref.~\cite{Titov:2002}. There are few features in this reaction off the proton that make it attractive for studying the role of nucleon resonances.  One is that the dominant contribution to the non-resonant amplitude is fairly known. At higher energies, the diffractive processes -- taken into account by Pomeron exchange -- dominate and, at lower energies, the $t$-channel pion exchange dominates \cite{Oh:2000zi}. In fact, this feature has been known since early 70ths \cite{ABBHHM:1968aa,Schilling:1968awz,Fraas:1972fj,Ballam:1972eq,Joos:1977tz,Friman:1995qm}. In Ref.~\cite{Sibirtsev:2003}, it is found that at intermediate energies below $W < 3.2$ GeV, Pomeron exchange is no longer sufficient to reproduce the data, and it has been suggested that $\pi$ and $f_2$ exchanges become the dominant contributions. Another feature is the isoscalar nature of the $\omega$ meson that filters out the isospin $I=3/2$ $\Delta$-resonances in the $s$-channel. Together, these features provide a great deal of simplification to the, otherwise a very complex problem of resonance extraction. The $\rho$ meson photoproduction also gives information on the resonances in the same mass region of $2$ GeV, since $\rho$ has a mass of about $775$ MeV which is close to the omega mass of $782.7$ MeV. However, the $\rho$ meson is an isovector meson and, as such, it excites not only $I=1/2$ but also $I=3/2$ resonances, which makes the analysis of resonance extraction more involved. In addition, unlike $\omega$, whose widths is about $8.5$ MeV, the width of $\rho$ is about $150$ MeV. This means that the effect of the width of $\rho$ in the final state has to be taken into account in the $\rho$ meson production reactions which may affect the quality of the associated experimental data.

Earlier data on $\omega$ photoproduction in the resonance energy region have low statistics and are scarce. They date from late 60ths to early 80ths and one in the late 90ths \cite{Crouch:1967, ABBHHM:1968aa, Eisenberg:1969, Ballam:1972eq, Eisenberg:1972, AHHM:1976, Barber:1984, Klein:1998}.  A new generation of data was reported only in 2003 by the SAPHIR Collaboration \cite{Barth:2003kv} with the first high-statistics cross section and spin-density-matrix elements (SDMEs) data in the center-of-mass energy range from the $\omega$-production threshold up to $W = 2.4$ GeV. In Ref.~\cite{Klein:2008aa}, the CBELSA/TAPS Collaboration reported the data on beam asymmetry in the near-threshold energy region. The CLAS Collaboration provided another high-statistics measurements of the differential cross section and SDMEs \cite{Williams:2009ab}. Many data were reported in the year 2015 in particular: new measurements of differential cross sections and SDMEs and the first measurements of the double polarization asymmetries, the beam-target-helicity asymmetries $E$ and $G$, were reported by the CBELSA/TAPS Collaboration \cite{Wilson:2015uoa, Eberhardt:2015lwa}; a new measurement of the differential cross section with full production-angle coverage was reported by A2 Collaboration at MAMI \cite{Strakovsky:2014wja}.  Also, the beam asymmetry $\Sigma$ has been measured by the GRAAL Collaboration \cite{Vegna:2013ccx}. More recently, the CLAS Collaboration reported the newest data of the photon beam asymmetry and a comparison with previous results were made \cite{Collins:2017vev}. The newest high-precision CLAS data on the spin observables $E$ \cite{Akbar:2017uuk} as well as on $\Sigma$, $T$ \cite{Roy:2017qwv} and $P'$, $F$, $H$ \cite{Roy:2018} have been just reported. The latter four observables, $T$, $P'$, $F$ and $H$, were measured for the very first time.

From the theoretical side, a number of studies of the reaction $\gamma p \to \omega p$ have been performed to learn about the reaction mechanisms in general and the role of nucleon resonances in particular.  Earlier studies focused more on learning about the basic features of this process such as the dominant non-resonant reaction mechanisms \cite{Schilling:1968awz,Friman:1995qm}. The formalism for vector meson photoproduction with polarized photon was given in an early work in Ref.~\cite{Schilling:1969um}. For a later work on spin information from the decay of vector meson in photoproduction, see Ref.~\cite{Kloet:1998}. The general aspect of the reaction based on symmetry considerations was investigated in Refs.~\cite{Pichowsky:1996,Savkli:1996}. The later works have concentrated on more specific aspects of the reaction such as the sensitivity of some of the spin-observables to certain details of the reaction dynamics \cite{Titov:1998,Sarantsev:2009} and the effects of the $\omega N$ final-state-interaction \cite{Oh:2002}. The role of the nucleon resonances in $\omega$ photoproduction has been investigated in Refs.~\cite{Zhao:2000tb,Oh:2000zi,Titov:2002iv} within tree-level effective Lagrangian approaches and in Refs.~\cite{Penner:2002md,Shklyar:2005} in a coupled-channel K-matrix approach. Also in Ref.~\cite{Paris:2009}, the $\omega$ photoproduction has been studied in a dynamical coupled channels approach. Only recently, with measurements of some of the spin-observables, in addition to high-statistics cross section data, the first partial-wave analyses have been performed \cite{Williams:2009aa,Denisenko:2016ugz}. Although these partial-wave analyses are limited,  improvements toward a more complete analyses are expected as the data base for this reaction increases, especially, with measurements of more independent spin-observables.

While all of the authors agree on the $\pi^0$ exchange in the $t$-channel playing a important role in the lower energy region and the diffractive processes dominance at higher energies, i.e., Pomeron exchange, they show discrepancies on various dominant resonance contributions. In the very first dedicated study of nucleon resonances in $\omega$ photoproduction, Zhao \cite{Zhao:2000tb} uses an effective quark model Lagrangian approach based on a quark model of Refs.~\cite{Zhao:1998a,Zhao:1998b}, and finds that the dominant resonances in his model are $N(1720)3/2^+$ and $N(1680)5/2^+$. The former is just at the nominal $\omega$ production threshold, while the latter is a below threshold resonance. Oh \textit{et al.} \cite{Oh:2000zi}, on the other hand, based on another quark model by Capstick and Roberts \cite{Capstick:1992,Capstick:1994}, claim that the dominant resonances are $N(1960)3/2^-$ and $N(1910)3/2^+$ in their calculation. The state $N(1910)3/2^+$ is a ``missing resonance" state predicted by the constituent quark model used to account for the configuration mixing. The calculation made by Titov and Lee \cite{Titov:2002iv}, where the resonance couplings are fixed from the empirical helicity amplitudes together with the vector meson dominance assumption, finds that the two most important resonance contributions to $\omega$ photoproduction come from $N(1680)5/2^+$ and $N(1520)3/2^-$ in the low energy region below $W =1.8$ GeV. Both resonances are below threshold resonances. The coupled-channel analysis of Giessen group \cite{Penner:2002md}, for energies $W \le 2$ GeV, finds a dominance of two resonances, $N(1710)1/2^+$ and $N(1900)3/2^+$, to the $\omega$ production mechanism. In a later analysis, including further data, the Giessen group \cite{Shklyar:2005} finds that while the $N(1680)5/2^+$ state only slightly influences the $\omega$ meson production in $\pi N$ scattering, its role is enhanced in $\omega$ photoproduction due to its relatively large electromagnetic coupling to the proton. In a dynamical coupled channel analysis, Paris \cite{Paris:2009} quotes the $S_{11}$, $D_{13}$ and $F_{15}$ partial waves as the dominant contributions to $\omega$ photoproduction, but the resonances are not extracted. Three states, $N(1700)3/2^+$, $N(1680)5/2^+$ and $N(2190)7/2^-$, are found to be dominant in the partial-wave analysis of Williams \textit{et al.} \cite{Williams:2009aa}, based on the high-statistics cross section and SDMEs data for energies up to $W=2.4$ GeV. The most recent (partial-wave) analysis of the $\omega$ photoproduction reaction to date has been done by Denisenko \textit{et al.} \cite{Denisenko:2016ugz} within the Bonn-Gatchina approach. It is, by far, the most complete analysis in the sense that it considers a large data base on pion and photo-induced reactions, including the recent $\omega$ photoproduction data for differential cross sections, several SDMEs, the beam asymmetry $\Sigma$, the normalized helicity difference $E$, and the correlation $G$ between linear photon and longitudinal target polarization. Here, twelve resonances, including the two nominally below-threshold resonances, $N(1700)3/2^-$ and $N(1710)1/2^+$, are found and the decay rates of these twelve resonances to
$N\omega$ were determined.

Most of the work mentioned above consider only the differential cross section data in their analyses due to the, then, lack of the data for more exclusive observables. By now, it is a well known fact that cross sections alone are far from imposing stringent constraints in the extraction of resonances. For this, measurements of spin polarization observables are of high significance and urgency. A complete experiment in vector meson photoproduction requires 24 independent observables to determine the 12 amplitudes \cite{Pichowsky:1996}. Thus, such an experiment is probably not feasible. Nevertheless, these more exclusive spin-observables are much more sensitive to the details of reaction dynamics in general and, therefore, are essential to help constrain the resonance content of existing models. Unlike for the scalar and pseudoscalar mesons, for vector mesons there is relatively a much larger number of independent spin observables. Thus, presumably they are more critical to help impose stringent constraints in the extraction of resonances. In fact, the sensitivity of some of the spin-observables in $\omega$ photoproduction to the reaction dynamics has been pointed out by various authors
\cite{Zhao:2000tb,Oh:2000zi,Titov:2002iv,Penner:2002md,Williams:2009aa,Denisenko:2016ugz}.

The purpose of the present study is to learn more about the role of resonances in $\omega$ meson photoproduction in the 2 GeV energy region based on the most recent high-precision unpolarized and polarized CLAS data. The analysis will be done within a tree-level effective Lagrangian approach, similar -- but not identical -- to that of Ref.~\cite{Oh:2000zi}. The strategy in the present work differs from that of Refs.~\cite{Zhao:2000tb,Oh:2000zi} in that, instead of using resonance couplings determined by quark models, we let the recent high-precision data to decide on the relevant resonances by fitting the resonance parameters to these data within our effective Lagrangian approach. Part of our theoretical results for double polarization observables $P'$, $F$ and $H$ have been published together with the experimental data from the CLAS Collaboration in Ref.~\cite{Roy:2018}. Here we report the details of our investigations and, in particular, we report our results for $P'$, $F$ and $H$ that are not shown in Ref.~\cite{Roy:2018} due to page limit and also the results for $\Sigma$, $T$, $E$, $d\sigma/d\Omega$ and SDMEs.

The paper is organized as follows. In Sec.~\ref{formalism}, the kinematics for the $\omega$ photoproduction reaction off the nucleon is defined as well as the effective Lagrangian densities for computing the corresponding amplitude and the formulas for calculating various observables are presented. Some brief comments on the gauge-invariant amplitude and the energy-dependent widths are also introduced in this section. Our results and discussion are presented in Sec.~\ref{results}. Finally a brief summary and conclusions are given in Sec.~\ref{summary}.

\section{Formalism}{\label{formalism}}

The reaction of interest in the present study is
\begin{equation}
\gamma(k) + p(p) \to \omega(q) + p(p'),   \label{eq:reaction}
\end{equation}
where the arguments $k$, $p$, $q$ and $p'$ indicate the four-momenta of the incoming photon, initial-state (target) proton, outgoing $\omega$ and final-state (recoil) proton, respectively. We also define the Mandelstam variables $t=(p-p^\prime)^2=(k-q)^2$, $s=(p+k)^2=(q+p')^2=W^2$ and $u=(p-q)^2=(p'-k)^2$.

\begin{figure}[tbp]
{\vglue 0.2cm}
\subfigure[~$t$ channel]{\includegraphics[width=0.45\columnwidth]{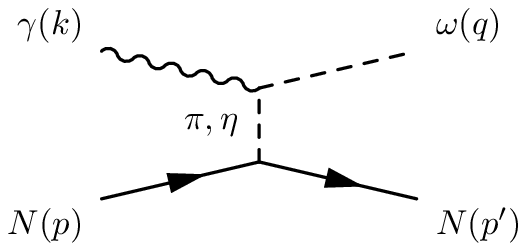}}
\subfigure[~$s$ channel]{\includegraphics[width=0.45\columnwidth]{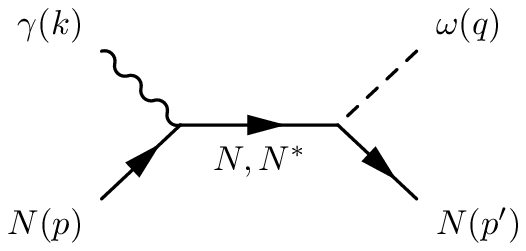}} \\[6pt]
\subfigure[~$u$ channel]{\includegraphics[width=0.45\columnwidth]{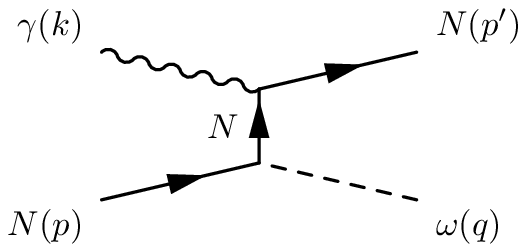}}
\subfigure[~Interaction current]{\includegraphics[width=0.45\columnwidth]{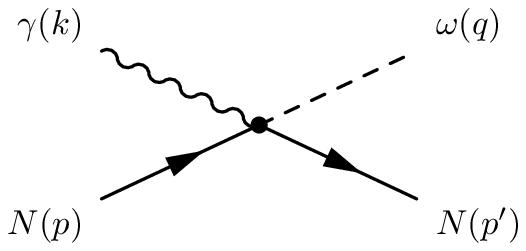}}
\caption{Feynman diagrams that define the present model for the $\gamma p \to \omega p$ reaction. Time proceeds from left to right.}
\label{fig:feynman}
\end{figure}

Following the field theoretical approach of Refs.~\cite{Haberzettl:1997,Haberzettl:2006bn,Huang:2012,Huang:2012xj}, the full amplitude for the present reaction can be expressed as
\begin{eqnarray}
M^{\nu\mu} = M^{\nu\mu}_s + M^{\nu\mu}_t + M^{\nu\mu}_u + M^{\nu\mu}_{\rm int},   \label{eq:amplitude}
\end{eqnarray}
with $\nu$ and $\mu$ being the Lorentz indices of the $\omega$ meson and photon, respectively. The first three terms $M^{\nu\mu}_s$, $M^{\nu\mu}_t$, and $M^{\nu\mu}_u$ stand for the contributions from the $s$-, $t$-, and $u$-channel diagrams, respectively. They arise from the photon attaching to the external particles  in the underlying three-point $NN\omega$ interaction vertex. The last term, $M^{\nu\mu}_{\rm int}$, stands for the interaction current which arises from the photon attaching to the internal structure of the $NN\omega$ interaction vertex. All four terms in Eq.~(\ref{eq:amplitude}) are diagrammatically depicted in Fig.~\ref{fig:feynman}. We refer the first term in Eq.~(\ref{eq:amplitude}), $M^{\nu\mu}_s$, as the resonant amplitude and the sum of the last three terms, $ M^{\nu\mu}_t + M^{\nu\mu}_u + M^{\nu\mu}_{\rm int}$, as the non-resonant amplitude.

The ($s$-channel) resonant amplitude consists of the nucleon and nucleon resonance contributions. As mentioned in the Introduction, previous studies of this reaction have shown that the non-resonant amplitude is dominated by the $t$-channel pion exchange at low energies and by the diffractive processes ($t$-channel Pomeron exchange) at higher energies. Since we are interested in the limited energy region below $W \sim 2.5$ GeV, the Pomeron exchange contribution can be safely ignored \cite{Oh:2000zi}. Thus, our non-resonant amplitude in the $t$-channel ($M^{\nu\mu}_t$) is given dominantly by the pion exchange contribution. We also include the $\eta$ meson exchange despite it's contribution being small since the $\omega\eta\gamma$ coupling  is known to be small. For the $u$-channel contribution to the non-resonant amplitude, we take into account the nucleonic current. We have checked and found that the $u$-channel resonance contributions are small enough to not affect the extracted resonance content. This is consistent with the observed behavior of the differential cross section data at backward angles.

The interaction current ($M^{\nu\mu}_{\rm int}$) contributes to the non-resonant amplitude very significantly, in general. However, for $\omega$ photoproduction, its contribution is expected to be relatively small, given that the non-resonant amplitude is dominated by the $t$-channel processes (pion plus Pomeron exchange). Hence, in the present work, the interaction current is taken into account in a minimal fashion by a phenomenological generalized contact current as specified below. This contact current is such that it preserves gauge invariance of the resulting total reaction amplitude (see, in particular, Refs.~\cite{Haberzettl:2006bn,Huang:2012}).

The Feynman diagrams that define the present model are shown in  Fig.~\ref{fig:feynman}. As mentioned above, they include (a) $t$-channel pseudo-scalar ($\pi$, $\eta$) meson  exchange current, (b) $s$-channel nucleonic ($N$) and resonance ($N^\ast$) currents, (c) $u$-channel nucleonic current and (d) the phenomenological contact current. In the following, we specify the effective Lagrangian densities and propagators required for constructing the reaction amplitude corresponding to the Feynman diagrams shown in Fig.~\ref{fig:feynman}.
The phenomenological (contact) interaction current is also specified.

\subsection{Effective Lagrangians} {\label{lagrangian}}

We calculate the $t$-channel mesonic  and  $u$-channel nucleonic currents by using the following effective Lagrangians:
\begin{align}
{\cal L}_{\omega\pi\gamma} & = \frac{e}{M_{\pi^0}} g_{\omega\pi\gamma} \epsilon^{\mu\nu\alpha\beta} \left(\partial_{\mu} A_\nu\right) \left(\partial_\alpha \pi^0 \right)  \omega_\beta, \label{eq:L-NNomega-1}  \\[3pt]
{\cal L}_{\omega\eta\gamma} & = \frac{e}{M_\eta}\, g_{\omega\eta\gamma} \epsilon^{\mu\nu\alpha\beta}\left(\partial_{\mu} A_\nu\right) \left(\partial_\alpha\eta\right)  \omega_\beta, \\[3pt]
{\cal L}_{NN\pi} & =- g_{NN\pi}\bar{N}\gamma_5 \left[\left(i \lambda  + \frac{1-\lambda}{2M_N} \slashed{\partial} \right)  {\boldsymbol \pi} \right] \cdot {\boldsymbol\tau} N, \\[3pt]
{\cal L}_{NN\eta} & =- g_{NN\eta}\bar{N}\gamma_5 \left[\left(i \lambda  + \frac{1-\lambda}{2M_N} \slashed{\partial} \right)  \eta\right] N, \\[3pt]
{\cal L}_{NN\gamma} &= -\,e \bar{N} \left[ \left( \hat{e} \gamma^\mu - \frac{ \hat{\kappa}_N} {2M_N}\sigma^{\mu \nu}\partial_\nu\right) A_\mu\right] N, \\[3pt]
{\cal L}_{NN\omega} & =-g_{NN\omega} \bar{N} \left[\left(\gamma_\mu - \frac{\kappa_\omega}{2M_N}\sigma_{\mu\nu}\partial^\nu \right) \omega^\mu\right] N,   \label{eq:L-NNomega}
\end{align}
where $\pi$, $A^\mu$, $\omega^\mu$ and $N$ denote the pion, photon, $\omega$ and nucleon field, respectively. The elementary charge unit is denoted by $e$, and $\hat{e} = (1+\tau_3)/2$ stands for the charge operator. The operator $\hat{\kappa}_N$ is equal to $\kappa_p\left(1+\tau_3\right)/2 + \kappa_n\left(1-\tau_3\right)/2$, with the anomalous magnetic moments $\kappa_p=1.793$ and $\kappa_n=-1.913$. In Eqs.~(\ref{eq:L-NNomega-1}) to (\ref{eq:L-NNomega}), $M_{\pi^0}$, $M_\eta$ and $M_N$ stand for the masses of $\pi^0$, $\eta$ and nucleon, respectively.

The coupling constant $g_{NN\pi}=13.46$, $g_{NN\eta}=4.76$, $g_{NN\omega}=11.76$ and $\kappa_\omega = 0$ are taken from a coupled-channel study of pion photoproduction of Ref.~\cite{Huang:2012}. The pseudoscalar-pseudovector mixing parameter $\lambda$ is set to $\lambda=0$ for both $\pi$ and $\eta$; for pion, this is demanded by chiral symmetry. The electromagnetic coupling constants $g_{\omega\pi\gamma}=0.32$ and $g_{\omega\eta\gamma}=0.25$ are also taken from the previous work of coupled-channel pion photoproduction \cite{Huang:2012}.

Following Refs.~\cite{Haberzettl:2006bn,Huang:2012xj}, the interaction current $M^{\nu\mu}_{\rm int}$ for $\gamma p \to \omega p$ is modeled in a minimal fashion by a generalized contact current
\begin{equation}
{\cal M}_{\rm int}^{\mu\nu} \approx {\cal M}_c^{\mu\nu} = \Gamma^\nu_{NN\omega}(q)C^\mu,  \label{eq:M_int}
\end{equation}
where $\Gamma^\nu_{NN\omega}$ stands for the $NN\omega$ vertex function given by the Lagrangian in Eq.~(\ref{eq:L-NNomega}),
\begin{equation}
\Gamma^\nu_{NN\omega}(q) = - i g_{NN\omega}\left[\gamma^{\nu}-i\frac{\kappa_\omega}{2M_N}\sigma^{\nu \alpha}q_{\alpha}\right].
\end{equation}
The auxiliary current $C^\mu$ in Eq.~(\ref{eq:M_int}) is given by
\begin{equation}
C^\mu=-e\frac{f_u-\hat F}{u-{p^\prime}^2}\left(2p^{\prime}-k\right)^\mu-e\frac{f_s-\hat F}{s-p^2}\left(2p+k\right)^\mu,
\end{equation}
with $f_u$ and $f_s$ denoting the phenomenological form factors, as specified later, in the $u$- and $s$-channel nucleonic current, respectively, and
\begin{equation}
\hat F=1- \hat{h} (1-f_s)(1-f_u).
\end{equation}
Here the parameter $\hat h$ may be an arbitrary (complex) function, $\hat h=\hat h(s, u, t)$, which, in general, is subject to crossing-symmetry constrains and must go to zero at high energies. The vanishing high-energy limit of $\hat{h}$ is necessary to prevent the ``violation of scaling behavior" \cite{Drell:1972}. Since we restricted ourselves to low energies in the present work, we set the fit parameter $\hat{h}$ as a constant, ${\hat h}=1$, for the sake of simplicity.

The $s$-channel resonance currents are constructed from the following Lagrangians. For the electromagnetic couplings, we have
\begin{align}
{\cal L}^{1/2\pm}_{RN\gamma}  =& \, e\frac{g^{(1)}_{RN\gamma}}{2M_N} {\bar R} \Gamma^{(\mp)}\sigma_{\mu\nu}\left(\partial^\nu A^\mu\right)N+ {\rm H.c.},  \\[3pt]
{\cal L}^{3/2\pm}_{RN\gamma}  =& -i e\frac{g^{(1)}_{RN\gamma}}{2M_N} {\bar R}_\mu\gamma_\nu\Gamma^{(\pm)}F^{\mu\nu}N   \nonumber \\
& + e\frac{g^{(2)}_{RN\gamma}}{\left(2M_N\right)^2} {\bar R}_\mu\Gamma^{(\pm)}F^{\mu\nu} \partial_{\nu} N + {\rm H.c.},  \\[3pt]
{\cal L}^{5/2\pm}_{RN\gamma}  =& \, e\frac{g^{(1)}_{RN\gamma}}{\left(2M_N\right)^2} {\bar R}_{\mu\alpha}\gamma_\nu\Gamma^{(\mp)} \left(\partial^\alpha F^{\mu\nu}\right) N  \nonumber \\
& \pm ie\frac{g^{(2)}_{RN\gamma}}{\left(2M_N\right)^3} {\bar R}_{\mu\alpha}\Gamma^{(\mp)} \left(\partial^\alpha F^{\mu\nu}\right)\partial_{\nu}N + {\rm H.c.},  \\[3pt]
{\cal L}^{7/2\pm}_{RN\gamma}  =& \, ie\frac{g^{(1)}_{RN\gamma}}{\left(2M_N\right)^3} {\bar R}_{\mu\alpha\beta}\gamma_\nu\Gamma^{(\pm)} \left(\partial^\alpha\partial^\beta F^{\mu\nu}\right) N  \nonumber \\
& - e\frac{g^{(2)}_{RN\gamma}}{\left(2M_N\right)^4} {\bar R}_{\mu\alpha\beta}\Gamma^{(\pm)}\left(\partial^\alpha\partial^\beta F^{\mu\nu}\right) \partial_{\nu}N + {\rm H.c.},
\end{align}
where $R$ designates the nucleon resonance, and the superscript of ${\cal L}_{RN\gamma}$ denotes the spin and parity of the resonance $R$. The coupling constants $g_{RN\gamma}^{(i)}$ $(i = 1, 2)$ are fit parameters. The notations $F^{\mu\nu}$ and $\Gamma^{(\pm)}$ are defined as
\begin{align}
F^{\mu\nu} & \equiv \partial^{\mu}A^{\nu}-\partial^{\nu}A^{\mu},  \\[3pt]
\Gamma^{(+)} &\equiv \gamma_5,  \quad \Gamma^{(-)}\equiv1.
\end{align}
For the hadronic couplings, we have
\begin{align}
{\cal L}^{1/2\pm}_{RN\omega} = &  -\frac{g_{RN\omega}}{2M_N} {\bar R} \Gamma^{(\mp)}  \left\{ \left[ \left( \frac{\gamma_\mu\partial^2}{M_R\mp M_N}\pm i\partial_\mu \right) \right.\right. \nonumber \\
& \left.\left. -\frac{f_{RN\omega}}{g_{RN\omega}} \sigma_{\mu\nu} \partial^\nu \right] \omega^\mu\right\} N + {\rm H.c.},  \\[3pt]
{\cal L}_{RN\omega}^{3/2\pm} =& - i\frac{g_{RN\omega}^{(1)}}{2M_N}\bar{R}_\mu \gamma_\nu \Gamma^{(\pm)}{\omega}^{\mu \nu} N  \nonumber \\
& + \frac{g_{RN\omega}^{(2)}}{\left(2M_N\right)^2} {\bar R}_\mu \Gamma^{(\pm)} \omega^{\mu \nu}\partial_\nu N \nonumber \\
& \mp \frac{g_{RN\omega}^{(3)}}{\left(2M_N\right)^2} {\bar R}_\mu \Gamma^{(\pm)}\left(\partial_\nu \omega^{\mu \nu}\right) N   + {\rm H.c.}, \\[3pt]
{\cal L}_{RN\omega}^{5/2\pm} = & \, \frac{g_{RN\omega}^{(1)}}{\left(2M_N\right)^2} {\bar R}_{\mu \alpha}\gamma_\nu \Gamma^{(\mp)}\left(\partial^{\alpha} \omega^{\mu \nu}\right) N   \nonumber \\
& \pm  i\frac{g_{RN\omega}^{(2)}}{\left(2M_N\right)^3} {\bar R}_{\mu \alpha} \Gamma^{(\mp)}\left(\partial^\alpha \omega^{\mu \nu}\right)\partial_\nu N  \nonumber \\
& \mp  i\frac{g_{RN\omega}^{(3)}}{\left(2M_N\right)^3} {\bar R}_{\mu \alpha} \Gamma^{(\mp)} \left(\partial^\alpha \partial_\nu \omega^{\mu \nu}\right) N  + {\rm H.c.}, \\[3pt]
 {\cal L}_{RN\omega}^{7/2\pm} =& \, i\frac{g_{RN\omega}^{(1)}}{\left(2M_N\right)^3} {\bar R}_{\mu \alpha \beta}\gamma_\nu \Gamma^{(\pm)} \left(\partial^{\alpha}\partial^{\beta} \omega^{\mu \nu}\right) N   \nonumber  \\
& -  \frac{g_{RN\omega}^{(2)}}{\left(2M_N\right)^4} {\bar R}_{\mu \alpha \beta} \Gamma^{(\pm)} \left(\partial^\alpha \partial^\beta \omega^{\mu \nu}\right) \partial_\nu N   \nonumber  \\
 & \pm \frac{g_{RN\omega}^{(3)}}{\left(2M_N\right)^4} {\bar R}_{\mu \alpha \beta} \Gamma^{(\pm)} \left(\partial^\alpha \partial^\beta \partial_\nu   \omega^{\mu \nu}\right) N  + {\rm H.c.},
\end{align}
where $\omega^{\mu\nu} \equiv \partial^\mu \omega^\nu - \partial^\nu \omega^\mu$. The parameters $g_{RN\omega}$, $f_{RN\omega}$ and $g_{RN\omega}^{(i)}(i=1,2,3)$ are fit parameters. We note that in tree-level calculations such as the present one, the results are only sensitive to the product of the electromagnetic and  hadronic couplings of the nucleon resonances.

\subsection{Resonance propagators}

For spin-$1/2$ resonance propagator, we use the ansatz
\begin{equation}
S_{1/2}(p) = \frac{i}{\slashed{p} - M_R + i \Gamma/2},
\end{equation}
where $M_R$, $\Gamma$ and $p$ are mass, width and four-momentum of the resonance $R$, respectively.

In accordance with Refs.~\cite{Behrends:1957,Fronsdal:1958,Zhu:1999}, the following prescriptions for the propagators of resonances with spin-$3/2$, -$5/2$ and -$7/2$ are adopted in the present work:
\begin{eqnarray}
S_{3/2}(p) &=&  \frac{i}{\slashed{p} - M_R + i \Gamma/2} \left( \tilde{g}_{\mu \nu} + \frac{1}{3} \tilde{\gamma}_\mu \tilde{\gamma}_\nu \right),  \\[3pt]
S_{5/2}(p) &=&  \frac{i}{\slashed{p} - M_R + i \Gamma/2} \,\bigg[ \, \frac{1}{2} \left(\tilde{g}_{\mu \alpha} \tilde{g}_{\nu \beta} + \tilde{g}_{\mu \beta} \tilde{g}_{\nu \alpha} \right)  \nonumber \\
&& -\, \frac{1}{5}\tilde{g}_{\mu \nu}\tilde{g}_{\alpha \beta}  + \frac{1}{10} \left(\tilde{g}_{\mu \alpha}\tilde{\gamma}_{\nu} \tilde{\gamma}_{\beta} + \tilde{g}_{\mu \beta}\tilde{\gamma}_{\nu} \tilde{\gamma}_{\alpha} \right. \nonumber \\
&& +\, \left. \tilde{g}_{\nu \alpha}\tilde{\gamma}_{\mu} \tilde{\gamma}_{\beta} +\tilde{g}_{\nu \beta}\tilde{\gamma}_{\mu} \tilde{\gamma}_{\alpha} \right) \bigg], \\[3pt]
S_{7/2}(p) &=&  \frac{i}{\slashed{p} - M_R + i \Gamma/2} \, \frac{1}{36}\sum_{P_{\mu} P_{\nu}} \bigg( \tilde{g}_{\mu_1 \nu_1}\tilde{g}_{\mu_2 \nu_2}\tilde{g}_{\mu_3 \nu_3} \nonumber \\
&& -\, \frac{3}{7}\tilde{g}_{\mu_1 \mu_2}\tilde{g}_{\nu_1 \nu_2}\tilde{g}_{\mu_3 \nu_3} + \frac{3}{7}\tilde{\gamma}_{\mu_1} \tilde{\gamma}_{\nu_1} \tilde{g}_{\mu_2 \nu_2}\tilde{g}_{\mu_3 \nu_3} \nonumber \\
&& -\, \frac{3}{35}\tilde{\gamma}_{\mu_1} \tilde{\gamma}_{\nu_1} \tilde{g}_{\mu_2 \mu_3}\tilde{g}_{\nu_2 \nu_3} \bigg),  \label{propagator-7hf}
\end{eqnarray}
where
\begin{eqnarray}
\tilde{g}_{\mu \nu} &=& -\, g_{\mu \nu} + \frac{p_{\mu} p_{\nu}}{M_R^2}, \\[3pt]
\tilde{\gamma}_{\mu} &=& \gamma^{\nu} \tilde{g}_{\nu \mu} = -\gamma_{\mu} + \frac{p_{\mu}\slashed{p}}{M_R^2},
\end{eqnarray}
and the summation over $P_\mu \left(P_\nu\right)$ in Eq.~(\ref{propagator-7hf}) goes over the $3!=6$ possible permutations of the indices $\mu_1\mu_2\mu_3\left(\nu_1\nu_2\nu_3\right)$.

The resonance width $\Gamma$ appearing in the resonance propagators given above is energy-dependent. We account for this dependence with an appropriate threshold behavior in our formalism. Explicitly, we write the width $\Gamma$ as a function of $W =\sqrt{s}$ in the form of
\begin{equation}
\Gamma(W)=\Gamma_R\left[\sum_{i=i}^N\beta_i\hat\Gamma_i(W)+\sum_{j=1}^{N_\gamma} \gamma_j \Gamma_{\gamma_j}(W)\right],
\end{equation}
where the sum over $i$ accounts for decays of the resonance into hadronic channels, and the sum over $j$ accounts for decays of the resonance into radiative channels. $\Gamma_R$ denotes the total static resonance width at $W=M_R$. The factors $\beta_i$ and $\gamma_j$ are, respectively, the hadronic and radiative decay branching ratios of the $i$-th resonance satisfying
\begin{equation}
\sum_{i=1}^N\beta_i+\sum_{j=1}^{N_\gamma}\gamma_j=1.   \label{eq:sumrule}
\end{equation}
Similar to Refs.~\cite{Nakayama:2005ts,Huang:2012xj}, we parameterize the width functions $\Gamma_i(W)$ and $\Gamma_{\gamma_j}(W)$ to provide the correct respective threshold behaviors. The details can be found in Ref.~\cite{Nakayama:2005ts}.

\subsection{Form Factors}

Each hadronic vertex obtained from the Lagrangians given in the previous subsection is accompanied with a phenomenological form factor to account for the composite nature of the hadrons. Following Ref.~\cite{Wang:2017tpe}, we take the form factor for intermediate baryons as
\begin{equation}
f_x(p_x^2) = \left(\frac{\Lambda_x^4}{\Lambda_x^4+\left(p_x^2-M_B^2\right)^2} \right)^2,  \label{eq:ff_B}
\end{equation}
where $p_x$ and $M_B$ denote the four-momentum and mass of the intermediate baryon either in the $s$- or $u$-channel as $x=s, u$. The cutoff parameters $\Lambda_x$ are treated as fit parameters.

For intermediate meson exchange, we take the form factor as \cite{Wang:2017tpe}
\begin{equation}
f_t(q_M^2) = \left(\frac{\Lambda_M^2-M_M^2}{\Lambda_M^2-q_M^2}\right)^2, \label{eq:ff_M}
\end{equation}
where $q_M$ and $M_M$ denote the four-momentum and mass of the intermediate meson, respectively. The cutoff parameters $\Lambda_M$ are treated as fit parameters.

\subsection{Observables}

We define a set of three mutually orthogonal unit vectors $\{\hat{x}, \hat{y}, \hat{z}\}$ in terms of the available momenta in the problem, i.e., the incident photon momentum ${\bf k}$ and the outgoing $\omega$-meson momentum ${\bf q}$:
\begin{equation}
\hat{z} \equiv \frac{{\bf k}}{| {\bf k} |},  \qquad   \hat{y} \equiv \frac{{\bf k} \times {\bf q}}{ |{\bf k} \times {\bf q} |}, \qquad \hat{x} \equiv \hat{y} \times \hat{z}.  \label{eq:refframe}
\end{equation}
Here, the boldface indicates the respective three-momentum.

In the $\gamma p \to \omega p$ center-of-mass (c.m.) frame, the invariant reaction amplitude $M^{\nu\mu}$, introduced in the previous subsection (cf.~Eq.~(\ref{eq:amplitude})), can be expressed in the c.m. helicity basis \cite{Jacob:1964}
\begin{equation}
T_{\lambda_\omega\lambda_f,\lambda_{\gamma}\lambda_i} \equiv \Braket{\textbf{q},\lambda_\omega; \textbf{p}_f,\lambda_f | M | \textbf{k},\lambda_{\gamma}; \textbf{p}_i,\lambda_i},     \label{eq:HelMtrx}
\end{equation}
where $\lambda_{\gamma}$, $\lambda_\omega$, $\lambda_i$ and $\lambda_f$ are the helicities of the proton, vector meson, initial and final state nucleons, respectively. The normalization of the above helicity amplitude is such that it is related to the differential cross section by
\begin{equation}
\frac{d\sigma}{d\Omega}= \frac{1}{64\pi^2 s}\frac{|\textbf{q}|}{|\textbf{k}|} \times  \frac{1}{4} \sum_{\lambda_{\gamma}\lambda_\omega \lambda_i \lambda_f} \left|{T_{{\lambda _\omega}{\lambda _f},{\lambda _\gamma }{\lambda _i}}}\right|^2. \label{eq:3-1}
\end{equation}

Together with cross sections, spin-polarization observables connect experiment with theory. They can be directly compared with theoretical calculations and hence give access to information about the reaction dynamics. In this work, we consider both the single and double spin-polarization observables involving the polarizations of the photon beam, target nucleon and recoil nucleon, in addition to the $\omega$-meson SDMEs with unpolarized target and recoil nucleons.

Following Ref.~\cite{Fasano:1992es}, we introduce the notation $\sigma(B, T; R, V)$ for the cross section $d\sigma/d\Omega$ where the arguments $(B, T; R, V)$ denote the polarizations of the photon beam ($B$), target proton ($T$), recoil proton ($R$) and produced $\omega$ meson ($V$), respectively. With this notation, the unpolarized differential cross section is
\begin{equation}
\frac{d\sigma}{d\Omega} = \sigma(U, U; U, U),   \label{eq:unpol-xsc}
\end{equation}
where $U$ denotes the unpolarized spin state.

For the single-polarization observables, photon beam asymmetry ($\Sigma$), target nucleon asymmetry ($T$), and recoil nucleon asymmetry ($P$), we have:
\begin{align}
\frac{d\sigma}{d\Omega} \Sigma & = \sigma(\bot, U; U, U) - \sigma(\|, U; U, U), \\[3pt]
\frac{d\sigma}{d\Omega} T & = \sigma(U, \bot; U, U) - \sigma(U, \|; U, U),  \\[3pt]
\frac{d\sigma}{d\Omega} P & = \sigma(U, U; \bot, U) - \sigma(U, U; \|, U),   \label{eq:spin-P}
\end{align}
where $\bot \left( \| \right)$ denotes the polarization state perpendicular (parallel) to the reaction plane.

For the double-polarization observables, we have
\begin{align}
\frac{d\sigma}{d\Omega} E  & =  2 \left[\sigma(r, -z, U, U) - \sigma(r, +z, U, U)\right], \\[3pt]
\frac{d\sigma}{d\Omega} G & = 2 \left[\sigma(\bot', +z, U, U) - \sigma(\bot', -z, U, U)\right], \\[3pt]
\frac{d\sigma}{d\Omega} F & = 2 \left[\sigma(r, +x, U, U) - \sigma(r, -x, U, U)\right], \\[3pt]
\frac{d\sigma}{d\Omega} H & = 2 \left[\sigma(\bot', -x, U, U) - \sigma(\bot', +x, U, U)\right],
\end{align}
where we label the helicity $+1$ circular polarization by $r$, and the helicity $-1$ circular polarization by $l$. $\bot'$ ($\|'$) stands for the polarization when the perpendicular $\bot$ (parallel $\|$) polarization is rotated clockwise about the $z$-axis by an angle $\phi=\pi/4$. $\pm z$ ($\pm x$) indicates the polarization in the $\pm \hat{z}$-direction ($\pm \hat{x}$-direction), respectively. We note that the above definitions of the single- and double-polarization observables coincides with those given in Ref.~\cite{Ronchen:2014}, except for an overall minus sign in the $E$- and $G$-observables.\footnote{Although different groups agree on the convention used for single-polarization observables, for double-polarization observables this is not the case and care must be taken in comparing these observables from different groups \cite{Sandorf:2012}.}

In Ref.~\cite{Roy:2018}, the CLAS Collaboration has measured the beam-target asymmetry $P'$,
\begin{align}
\frac{d\sigma}{d\Omega} P' = &\left[\sigma(\bot, +y, U, U) - \sigma(\bot, -y, U, U) \right. \nonumber \\
& \left. - \, \sigma(\parallel, +y, U, U) + \sigma(\parallel, -y, U, U)\right],   \label{eq:spin-P'}
\end{align}
which is called $P$ in Ref.~\cite{Roy:2018} but is different from the usually used one defined in Eq.~(\ref{eq:spin-P}) for the single spin observable, recoil nucleon asymmetry $(P)$.\footnote{Only for pseudo-scalar meson photoproduction, the $P'$ defined in Eq.~(\ref{eq:spin-P'}) is the same to the recoil asymmetry $P$ defined in Eq.~(\ref{eq:spin-P}).} In the present work, we shall use the symbol $P'$ instead of $P$ for the beam-target asymmetry observable defined in Eq.~(\ref{eq:spin-P'}).

The SDMEs, as the interference between the independent helicity amplitudes, can be measured in the final state $\omega$ decay distribution.  Explicitly, they are given by \cite{Schilling:1969um}
\begin{align}
\rho^0_{\lambda_\omega \lambda'_\omega} & = \frac{1}{{2N}}\sum\limits_{{\lambda_\lambda} {\lambda_f} {\lambda _i}} {{T_{{\lambda_\omega} {\lambda_f}, {\lambda_\gamma} {\lambda_i}}} T^*_{{\lambda'_\omega} {\lambda_f}, {\lambda_\gamma} {\lambda _i}}}, \\[3pt]
\rho^1_{{\lambda_\omega}{\lambda'_\omega}} & = \frac{1}{{2N}}\sum\limits_{{\lambda_\lambda} {\lambda_f} {\lambda_i}} {{T_{{\lambda_\omega} {\lambda_f}, -{\lambda_\gamma} {\lambda_i}}} T^*_{{\lambda'_\omega} {\lambda_f}, {\lambda_\gamma} {\lambda _i}}},  \\[3pt]
\rho^2_{{\lambda_\omega}{\lambda'_\omega}} & = \frac{i}{{2N}}\sum\limits_{{\lambda_\lambda} {\lambda_f} {\lambda_i}} \lambda_\gamma {{T_{{\lambda_\omega} {\lambda_f}, -{\lambda_\gamma} {\lambda_i}}} T^*_{{\lambda'_\omega}{\lambda_f}, {\lambda_\gamma} {\lambda_i}}}, \\[3pt]
\rho^3_{{\lambda _\omega} {\lambda'_\omega}} & = \frac{1}{{2N}}\sum\limits_{{\lambda_\lambda} {\lambda_f} {\lambda_i}} \lambda_\gamma {{T_{{\lambda_\omega} {\lambda_f}, {\lambda_\gamma } {\lambda_i}}} T^*_{{\lambda'_\omega} {\lambda_f}, {\lambda_\gamma} {\lambda_i}}}.
\end{align}
Here $\lambda's$ denote the helicities of the respective particles as given in Eq.~(\ref{eq:HelMtrx}), and $N$ is the normalization factor given by
\begin{equation}
N \equiv \frac{1}{2}\sum\limits_{{\lambda_\omega }{\lambda_f}{\lambda _i}{\lambda_\gamma }} \left| {T_{{\lambda_\omega}{\lambda_f}, {\lambda_\gamma }{\lambda_i}}} \right|^2.
\end{equation}

Because the particle's decay distributions in general, and the $\omega$ decay distribution in particular, are measured in the decaying particle's rest frame, the SDMEs are usually presented also in this frame. More specifically, there are three reference frames of common use, which differ from each other by the choice in the quantization axis only.  In the helicity frame, the three mutually orthogonal axes are given by
\begin{equation}
\hat{z}' \equiv \frac{{\bf q}}{|{\bf q}|},    \qquad     \hat{y}' \equiv \frac{{\bf k} \times {\bf q}}{ \left|{\bf k} \times {\bf q} \right|}, \qquad    \hat{x}' \equiv \hat{y}' \times \hat{z}'.  \label{eq:HelFrame}
\end{equation}
In the Gottfried-Jackson frame, the choice of the orthogonal axes is the same to that of Eq.~(\ref{eq:refframe}). Thus this frame is simply the c.m. frame boosted by to the $\omega$ rest frame.
In the Adair frame, the $\hat{z}'$-axis equals the direction of the photon momentum as measured in the c.m. frame.
The three frames defined above are related by a simply rotation about the $\hat{y}'$-axis.
For forward-produced $\omega$ meson, all three frames coincide.

\section{Numerical results and discussions} {\label{results}}

\begin{table*}[htbp]
\caption{\label{table:decay channel} The resonance decay channels and the corresponding branching ratios (in $\%$) considered in the present model. Most of the hadronic decay branching ratios in bold font denote the centroid values of the dominant decay modes quoted by PDG \cite{PDG:2018}. The electromagnetic branching ratios  have been also fixed at the centroid values quoted by PDG, except for the $N(1860)5/2^+$ resonance, of which no information on $N\gamma$ branching ratio is given in PDG. The other branching ratios in normal font are determined by the present fit to the $\omega$ photoproduction data.}
\renewcommand{\arraystretch}{1.2}
\begin{tabular*}{\textwidth}{@{\extracolsep\fill}lccccccc}
\hline\hline
   & $N(1520)3/2^-$ & $N(1700)3/2^-$ & $N(1720)3/2^+$ & $N(1860)5/2^+$ & $N(1875)3/2^-$ & $N(1895)1/2^-$ & $N(2060)5/2^-$    \\ \hline
$N\pi$      & $\bm{60}$   & $\bm{12}$     & $\bm{11}$     & $\bm{12}$     & $\bm{7}$       & $\bm{10}$      & $\bm{10}$      \\
$N\pi\pi$   & $\bm{11.58}$ &              & $19.31\pm0.01$ & $18.74\pm0.02$ & $0.00\pm0.01$  &               & $39.31\pm0.01$ \\
$\Delta\pi$ & $\bm{28}$   & $\bm{43}$     & $\bm{68}$  &              & $\bm{22}$     &               & $\bm{7}$        \\
$N\rho$     &            & $\bm{23}$     &              &              &               &               &                \\
$N\eta^\prime$&          &              &              &              &               & $\bm{21}$      &                \\
$N\sigma$   &            &              &              & $\bm{41}$    & $\bm{45}$     &               &                 \\
$N\eta$     &            &              &              &              &               & $\bm{23}$      &                 \\
$\Lambda K$ &            &              &              &              &               & $\bm{15}$      &                 \\
$\Sigma K$  &            &              &              &              &               & $\bm{11}$      &                \\
$N(1440)\pi$&            & $\bm{7}$      &              &              &               &               & $\bm{9}$       \\
$N(1520)\pi$&            &              &              &              &               &               & $\bm{15}$       \\
$N(1680)\pi$&            &              &              &              &               &               & $\bm{15}$       \\
$N\omega$   &     & $14.97\pm0.01$ & $1.54\pm0.01$ & $28.21\pm0.02$ & $25.99\pm0.01$ & $19.97\pm0.01$ & $4.58\pm0.01$      \\
                     &      & $[10 - 34]$      &  $[12 - 40]$      &        & $[15 - 25]$              & $[16 - 40]$ & $[1 - 7]$     \\
$N\gamma$ & $\bm{0.42}$ & $\bm{0.03}$ & $\bm{0.15}$  & $\bm{0.05}$   & $\bm{0.013}$   & $\bm{0.035}$   & $\bm{0.11}$      \\
                     & $[0.31 - 0.52]$       & $[0.01 - 0.05]$            & $[0.05 - 0.25]$    &         & $[0.001 - 0.025]$              & $[0.01 - 0.06]$  &  $[0.03 - 0.19]$     \\
\hline\hline
\end{tabular*}
\end{table*}

\begin{table*}[htbp]
\caption{\label{table:constants} Fitted values for the product of the coupling constants and the phase factor for each resonance. Here for resonances with spin $1/2$, $g^{(1)}_{RN\omega}$ and $g^{(2)}_{RN\omega}$ stand for $g_{RN\omega}$ and $f_{RN\omega}$, respectively. }
\renewcommand{\arraystretch}{1.2}
\begin{tabular*}{\textwidth}{@{\extracolsep\fill}lcccccccc}
\hline \hline
                                 & $g^{(1)}_{RN\gamma}*g^{(1)}_{RN\omega}$  & $g^{(2)}_{RN\gamma}*g^{(1)}_{RN\omega}$  & $g^{(1)}_{RN\gamma}*g^{(2)}_{RN\omega}$  & $g^{(2)}_{RN\gamma}*g^{(2)}_{RN\omega}$ & $g^{(1)}_{RN\gamma}*g^{(3)}_{RN\omega}$  & $g^{(2)}_{RN\gamma}*g^{(3)}_{RN\omega}$  & $\phi_R/\pi$    \\ \hline
$N(1520)3/2^-$  & $162.35\pm0.04$  & $-175.17\pm0.04$  & $-342.93\pm0.06$ & $370.01\pm0.17$ & $8.19\pm0.04$ & $-8.84\pm0.04$  & $1.293\pm0.001$  \\
$N(1700)3/2^-$  & $78.64\pm0.25$   & $-113.48\pm0.40$  & $-91.50\pm0.38$ & $132.03\pm0.59$  & $-24.11\pm0.18$ & $34.79\pm0.27$ & $0.042\pm0.001$ \\
$N(1720)3/2^+$  & $-6.86\pm0.01$  & $7.44\pm0.01$  & $12.18\pm0.01$  & $-13.22\pm0.01$ &  $-0.81\pm0.01$   &  $0.87\pm0.01$   &  $1.939\pm0.001$  \\
$N(1860)5/2^+$  & $147.21\pm0.20$  & $97.62\pm0.17$ & $-98.23\pm0.21$  &  $-65.14\pm0.05$ & $75.00\pm0.02$ & $49.73\pm0.16$   & $0.113\pm0.001$  \\
$N(1875)3/2^-$  & $15.18\pm0.03$  & $-13.59\pm0.04$ & $-21.02\pm0.04$  & $18.81\pm0.05$ & $-5.16\pm0.01$ & $4.62\pm0.02$    & $0.763\pm0.001$   \\
$N(1895)1/2^-$  & $1.64\pm0.01$  &  & $-0.08\pm0.01$                &          &        &  & $1.432\pm0.001$        \\
$N(2060)5/2^-$ & $-14.63\pm0.02$ & $0.20\pm0.03$  & $143.06\pm0.05$  & $-1.92\pm0.25$ & $123.31\pm0.04$ & $-1.65\pm0.21$  & $0.142\pm0.001$ \\ 
 \hline\hline
\end{tabular*}
\end{table*}

\begin{table}[htbp]
\caption{\label{table:cutoff} Fitted values for the cutoff parameters $\Lambda_t$, $\Lambda_s$ and $\Lambda_u$ of the form factors in the $t$-, $s$-, and $u$-channel non-resonant amplitudes.}
\renewcommand{\arraystretch}{1.2}
\begin{tabular*}{\columnwidth}{@{\extracolsep\fill}ccc}
\hline \hline
     $\Lambda_t$ [MeV]   & $\Lambda_s$ [MeV]   & $\Lambda_u$  [MeV]   \\ \hline
       $700\pm 6$          & $1194\pm 16$         & $500\pm 58$        \\
\hline\hline
\end{tabular*}
\end{table}

\begin{table}[htbp]
\caption{\label{table:mass and width} Fit values for the resonances mass $M_R$, width $\Gamma_R$ and cutoff parameter $\Lambda_R$. The quantities in square brackets are the quoted ranges in PDG \cite{PDG:2018}. }
\renewcommand{\arraystretch}{1.2}
\begin{tabular*}{\columnwidth}{@{\extracolsep\fill}ccccc}
\hline\hline
     &  status &  $M_R$ [MeV]  &  $\Gamma_R$ [MeV]  &  $\Lambda_R$ [MeV]     \\
\hline
$N(1520)3/2^-$   &  ****  &  $1508\pm10$  &  $98\pm9$  &  $1116\pm7$       \\
                          &             & $[1510\sim1520]$  &  $[100\sim120]$   &       \\   \hline
$N(1700)3/2^-$   &  ***   &  $1721\pm1$   &  $295\pm16$  &  $1200\pm12$          \\
                          &            &  $[1650\sim1800]$  & $[100\sim300]$  &         \\   \hline
$N(1720)3/2^+$   &  ****  & $1736\pm2$    &  $140\pm3$  &  $1650\pm13$       \\
                          &            &  $[1680\sim1750]$  &  $[150\sim400]$  &        \\  \hline
$N(1860)5/2^+$   &  **    &  $1800\pm3$   & $201\pm10$   &  $991\pm18$          \\  \hline
$N(1875)3/2^-$   &  ***  & $1943\pm10$     & $205\pm4$   & $1500\pm32$          \\
                         &            &  $[1850\sim1920]$  & $[120\sim250]$   &        \\ \hline
$N(1895)1/2^-$   &  ****   & $1953\pm3$     & $272\pm23$   & $1500\pm67$          \\ 
                         &            &  $[1870\sim1920]$  & $[80\sim200]$   &        \\ \hline
$N(2060)5/2^-$  &  ***   & $2030\pm3$    & $360\pm23$   &  $1212\pm38$       \\ 
                         &            &  $[2030\sim2220]$  & $[300\sim450]$   &        \\ 
\hline\hline
\end{tabular*}
\end{table}

High-precision experimental data for cross sections and a number of spin observables in $\gamma p \to \omega p$ have been reported by different groups (SAPHIR, CLAS, A2 and CBELSA/TAPS). Some of the data sets from these groups, however, are inconsistent with each other. This is the case, e.g., with the cross section data sets. The SAPHIR \cite{Barth:2003kv}, CLAS \cite{Williams:2009ab} and A2 \cite{Strakovsky:2014wja} Collaborations' data sets are in fairly good agreement overall with each other in the overlapping energy regions but there are some noticeable discrepancies, especially, close to threshold energies, where the SAPHIR and CLAS data are poorer than the A2 data. The CBELSA/TAPS cross section data \cite{Wilson:2015uoa} show a clear discrepancy  (which increases with energy and becomes clear for energies above $W \sim 2$ GeV or so) with both the SAPHIR and CLAS data. The SAPHIR data seem systematically lower than both the CLAS and CBELSA/TAPS  data for backward $\omega$ production angles. Despite a considerable effort has been made to resolve this issue, in particular, between the CBELSA/TAPS and CLAS Collaborations, where the discrepancy appears to be almost a linear energy-dependent normalization factor, the nature of the discrepancy remains unclear  \cite{Wilson:2015uoa}. This situation makes difficulty to include the data sets from these groups in a single analysis.  In particular, there is no valid reason to discard any one of these data sets in favor of others. In the present work, we chose to include in our analysis the CLAS  data only because they have, in addition to the differential cross sections, the newest and largest number of independent spin observables data with higher accuracies than those from other groups. In addition to the differential cross sections and SDMEs data \cite{Williams:2009ab}, the CLAS Collaboration has reported the newest data of $\Sigma$, $T$ \cite{Roy:2017qwv}, $E$ \cite{Akbar:2017uuk}, and, just recently, $P'$, $F$, $H$ \cite{Roy:2018}; the data for $T$, $P'$, $F$, and $H$ were measured for the very first time. Later, we consider the A2 and CBELSA/TAPS data separately.

The major objective of the present work is to extract the information on the nucleon resonances involved in the reaction $\gamma p \to \omega p$ based on an effective Lagrangian approach as described in Sec.~\ref{formalism}. For this purpose, in addition to the cross section data, we include all available new data on spin observables as mentioned above to constrain the model parameters. This is the main difference that distinguishes the present work from earlier analyses.  We confine our analysis in the energy region from threshold  up to $\sim 2.25$ GeV. The strategy employed in this work is to consider a minimum number of resonances to achieve an acceptable description of these data. We consider the resonances listed in PDG \cite{PDG:2018} in the energy range of $W = 1.7 \sim 2.0$ GeV. Specifically, we find that the data can be satisfactorily described by including the set of following resonances: $N(1700)3/2^-$, $N(1720)3/2^+$, $N(1860)5/2^+$, $N(1875)3/2^-$, and $N(1895)1/2^-$ and $N(2060)5/2^-$, in addition to the sub-threshold $N(1520)3/2^-$. We mention that we have tried all other possible combinations of $7$ or less resonances before arriving to our solution for the set of resonances. Adding one more resonance to the achieved solution will not improve the fitting quality considerably, and we thus postpone such attempts until more spin observable data become available.

The model parameters associated with resonance are the resonance mass $M_R$, total decay width $\Gamma_R$ at $W=M_R$, decay branching ratios $\beta_i$,  the electromagnetic and hadronic coupling constants, $g^{(1,2)}_{RN\gamma}$, $g_{RN\omega}$, $f_{RN\omega}$ and $g^{(1,2,3)}_{RN\omega}$, in addition to the cutoff parameter $\Lambda_R$ in the form factor. They are free parameters to be fitted to reproduce the considered data. Whenever available, the dominant hadronic decay branching ratios are taken from PDG \cite{PDG:2018}. We then introduce one additional (effective) decay channel to satisfy the sum rule given by Eq.~(\ref{eq:sumrule}). Of course, for resonances above the $N\omega$-threshold, we have in addition the $N\omega$ branching ratio which is determined from the corresponding fitted $RN\omega$ coupling constants. The radiative decay branching ratios are also taken from PDG. In the present work we allow both the electromagnetic and hadronic coupling constants to be complex. With the observation that in our tree-level effective Lagrangian approach, the results are sensitive only to the product of the electromagnetic and hadronic vertex functions, we introduce a common complex phase $e^{{\rm i}\phi_R}$ to the product of the coupling constants, $g^{(i)}_{RN\gamma} g^{(j)}_{RN\omega}$, for a given resonance $R$ to account for the complex nature of these coupling constants. All the parameters of the non-resonant $u$-, $t$- and (nucleon) $s$-channel amplitudes have been calculated or fixed from independent sources as described in Sec.~\ref{formalism}, except for the cutoff parameters in the form factors, $\Lambda_t$, $\Lambda_u$ and $\Lambda_s$, which are treated as fit parameters.

\begin{figure*}[htbp]
\includegraphics[width=0.75\textwidth]{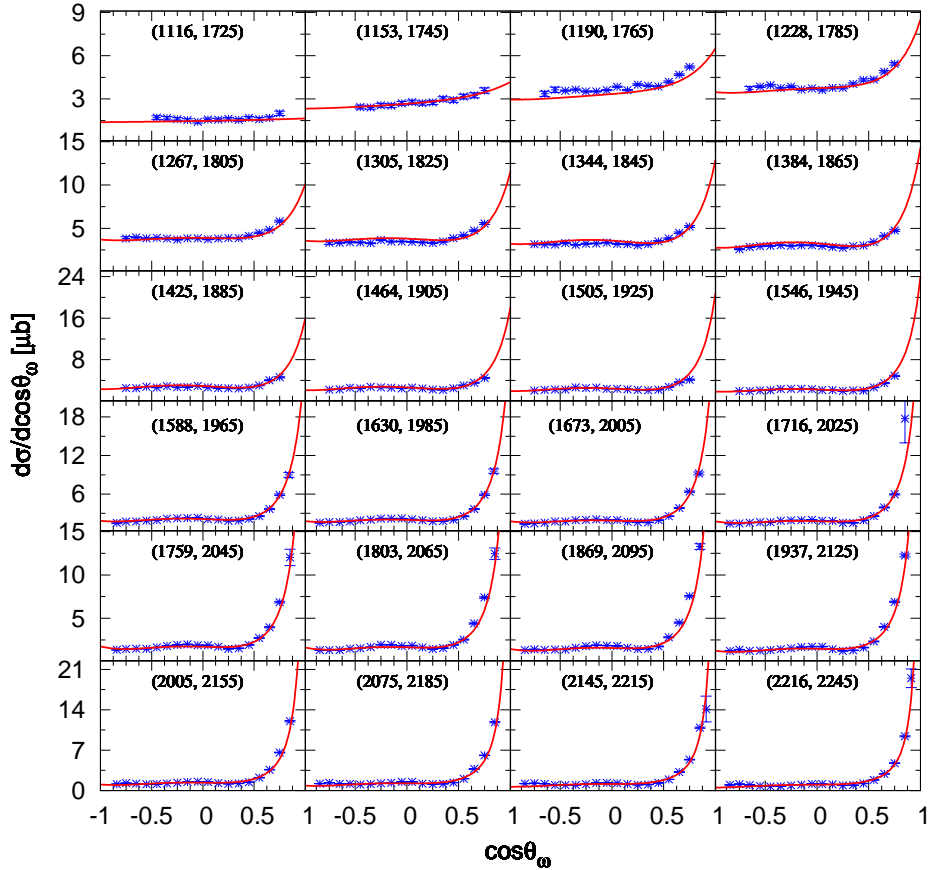}
\caption{Differential cross section for $\gamma p \rightarrow \omega p$ as a function of $\cos\theta_\omega$ in the center-of-mass frame at energy from near threshold region to $2.245$ GeV. The numbers in parentheses denote the photon incident energy (left number) and the corresponding center-of-mass energy of the system (right number), in MeV. Data are taken from CLAS Collaboration \cite{Williams:2009ab}. }
\label{pic:dsig}
\end{figure*}

\begin{figure*}[htbp]
\includegraphics[width=0.75\textwidth]{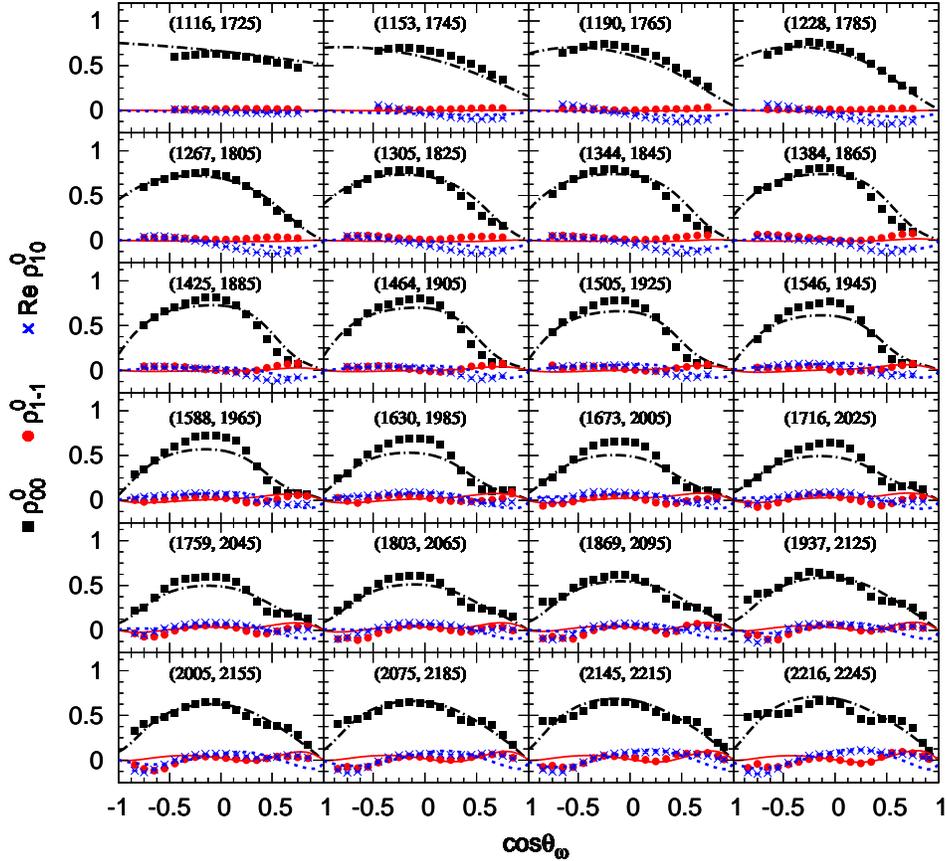}
\caption{Spin Density Matrix Elements in the Adair frame for $\gamma p \rightarrow \omega p$ as a function of $\cos\theta_\omega$ in the center-of-mass frame at energy from near threshold region to $2.245$ GeV. The black squares denote $\rho^0_{00}$, the red circles denote $\rho^0_{1-1}$ and the blue crosses denote Re\,$\rho^0_{10}$, respectively. Numbers in parentheses are defined in Fig.~\ref{pic:dsig} caption. Data are taken from CLAS Collaboration \cite{Williams:2009ab}.}
\label{pic:rho}
\end{figure*}

\begin{figure*}[htbp]
\includegraphics[width=0.75\textwidth]{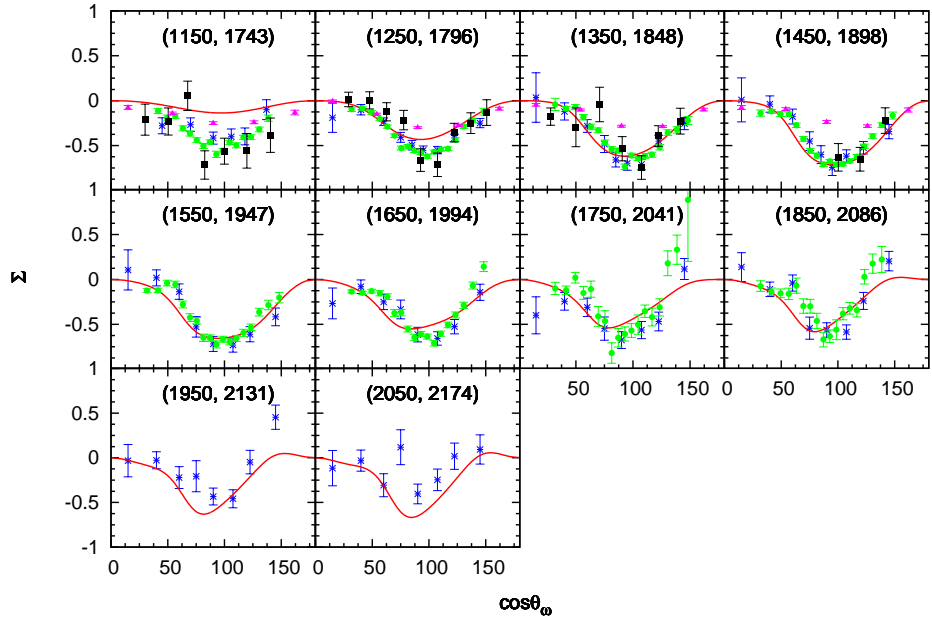}
\caption{Beam asymmetry $\Sigma$ for $\gamma p \rightarrow \omega p$ as a function of $\cos\theta_\omega$ in the center-of-mass frame. Numbers in parentheses are defined in Fig.~\ref{pic:dsig} caption. Data are taken from CLAS Collaboration \cite{Roy:2018} (star), \cite{Collins:2017vev} (circle), CBELSA/TAPS Collaboration \cite{Denisenko:2016ugz} (square), GRAAL Collaboration \cite{Vegna:2013ccx} (triangle). Only the data from Ref.~\cite{Roy:2018} are included in the fit. }
\label{pic:sigma}
\end{figure*}

\begin{figure*}[htbp]
\includegraphics[width=0.75\textwidth]{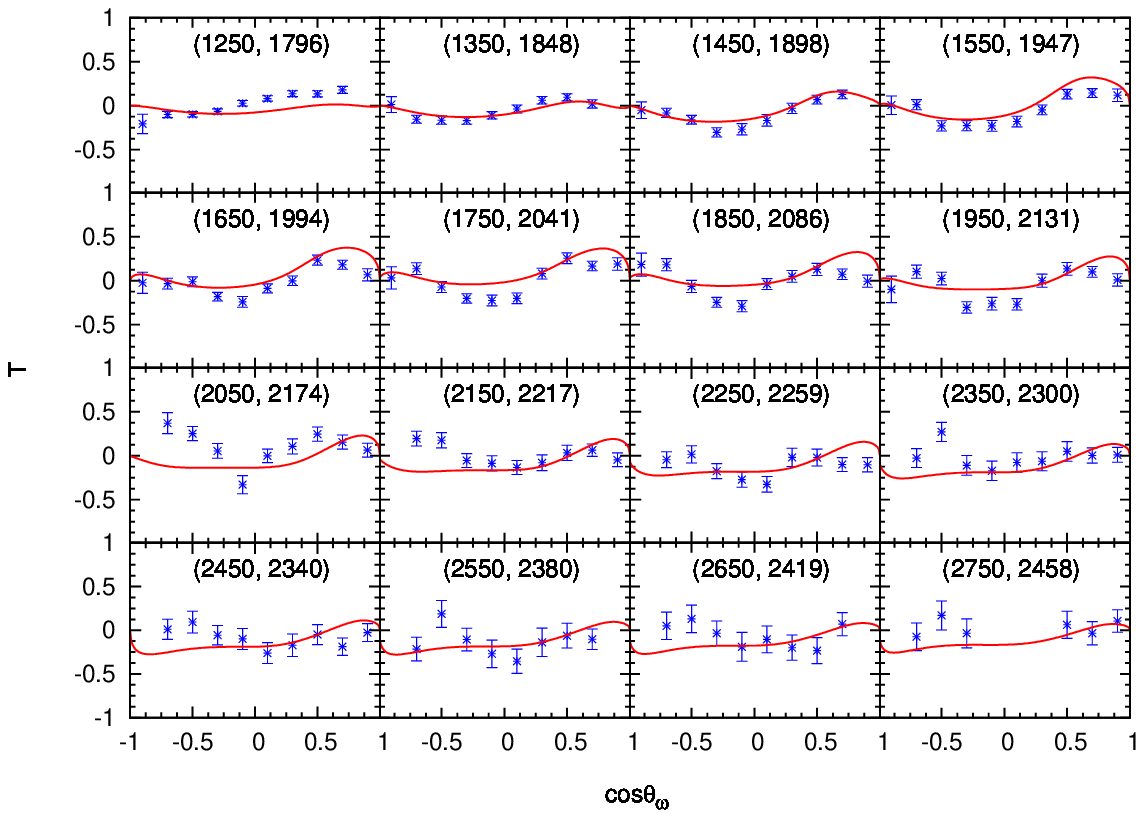}
\caption{Target asymmetry $T$ for $\gamma p \rightarrow \omega p$ as a function of $\cos\theta_\omega$ in the center-of-mass frame. Numbers in parentheses are defined in Fig.~\ref{pic:dsig} caption. Data are taken from CLAS Collaboration \cite{Roy:2018}. Only the data with center-of-mass energy smaller than $2.245$ GeV are included in the fit.  }
\label{pic:t}
\end{figure*}

\begin{figure*}[htbp]
\includegraphics[width=0.75\textwidth]{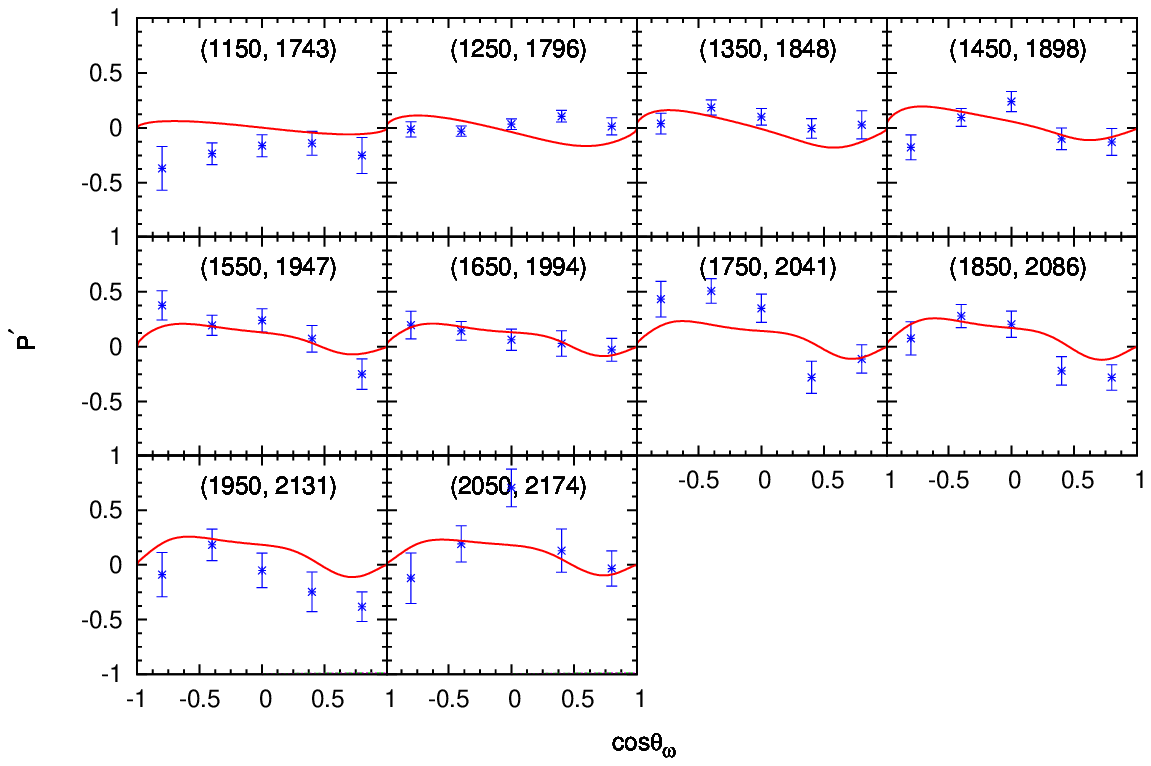}
\caption{Beam-target asymmetry $P'$ for $\gamma p \rightarrow \omega p$ as a function of $\cos\theta_\omega$ in the center-of-mass frame. Numbers in parentheses are defined in Fig.~\ref{pic:dsig} caption. Data are taken from CLAS Collaboration \cite{Roy:2018}. }
\label{pic:p}
\end{figure*}

\begin{figure*}[htbp]
\includegraphics[width=0.75\textwidth]{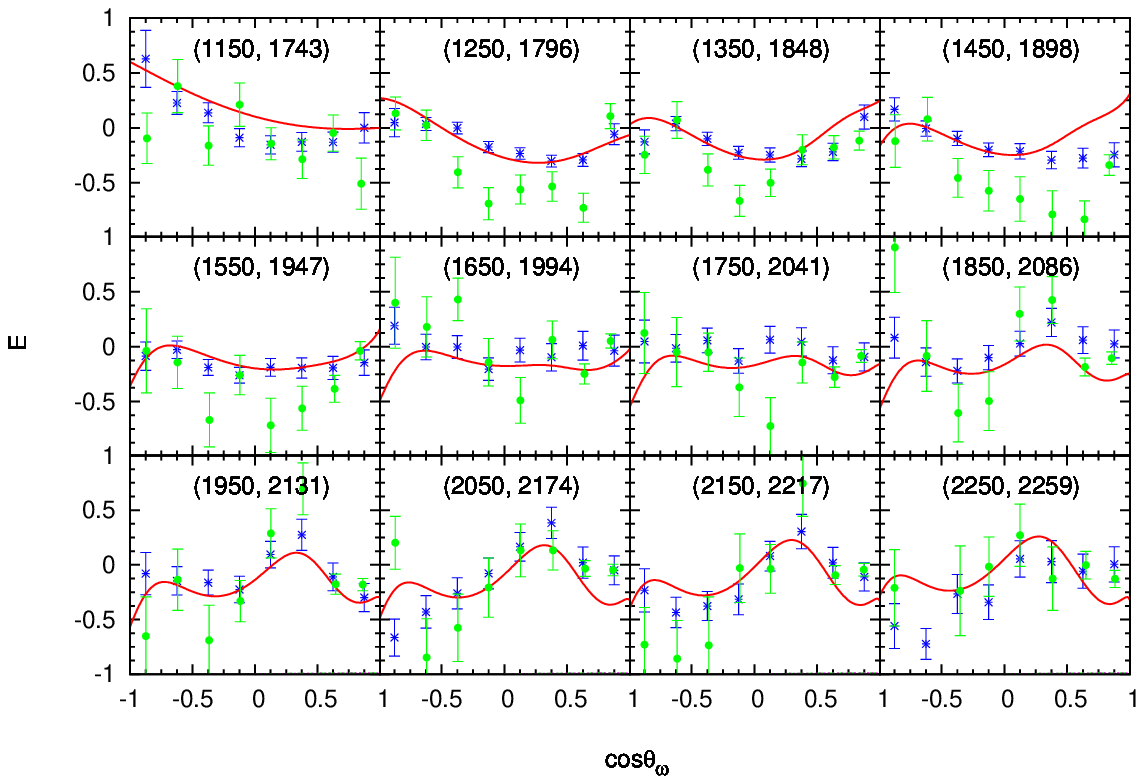}
\caption{Beam-target asymmetry $E$ for $\gamma p \rightarrow \omega p$ as a function of $\cos\theta_\omega$ in the center-of-mass frame. Numbers in parentheses are defined in Fig.~\ref{pic:dsig} caption. Data are taken from CLAS Collaboration \cite{Akbar:2017uuk} (star) and CBELSA/TAPS Collaboration \cite{Denisenko:2016ugz} (circle). Only the data from Ref.~\cite{Akbar:2017uuk} with center-of-mass energy smaller than $2.245$ GeV are included in the fit. }
\label{pic:e}
\end{figure*}

\begin{figure*}[htbp]
\includegraphics[width=0.75\textwidth]{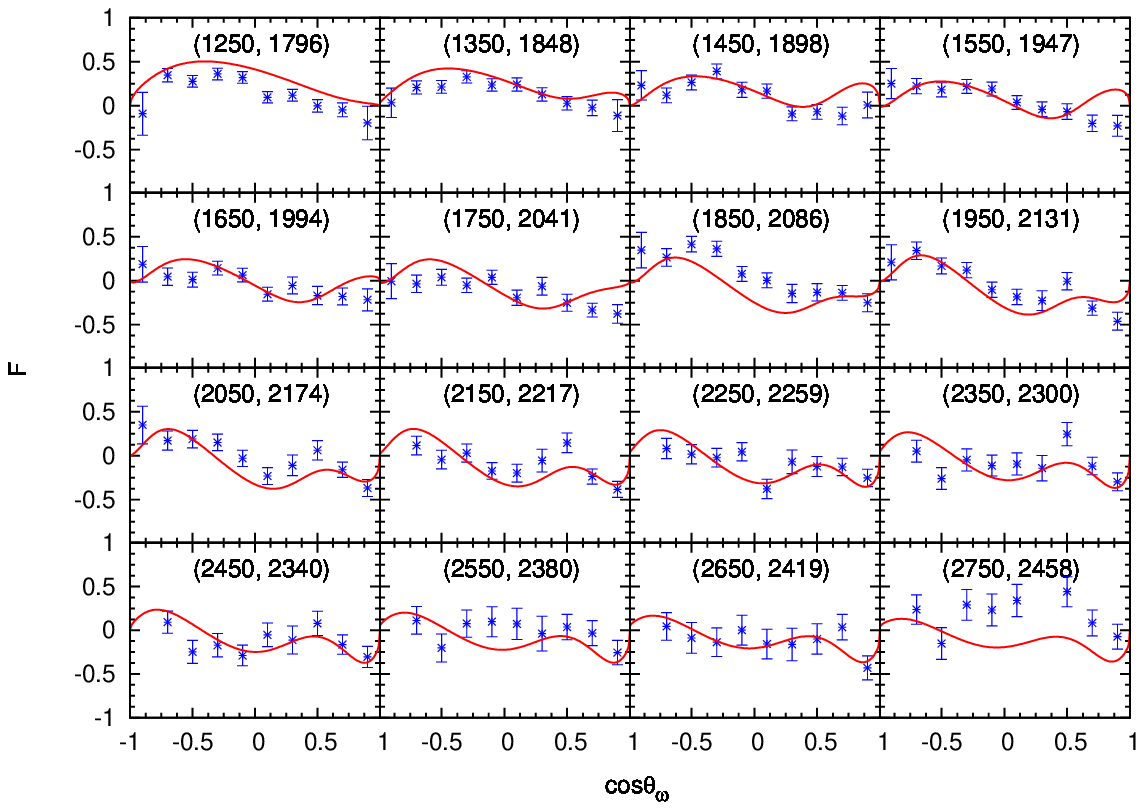}
\caption{Beam-target asymmetry $F$ for $\gamma p \rightarrow \omega p$ as a function of $\cos\theta_\omega$ in the center-of-mass frame. Numbers in parentheses are defined in Fig.~\ref{pic:dsig} caption. Data are taken from CLAS Collaboration \cite{Roy:2018}. Only the data with center-of-mass energy smaller than $2.245$ GeV are included in the fit.}
\label{pic:f}
\end{figure*}

\begin{figure*}[htbp]
\includegraphics[width=0.75\textwidth]{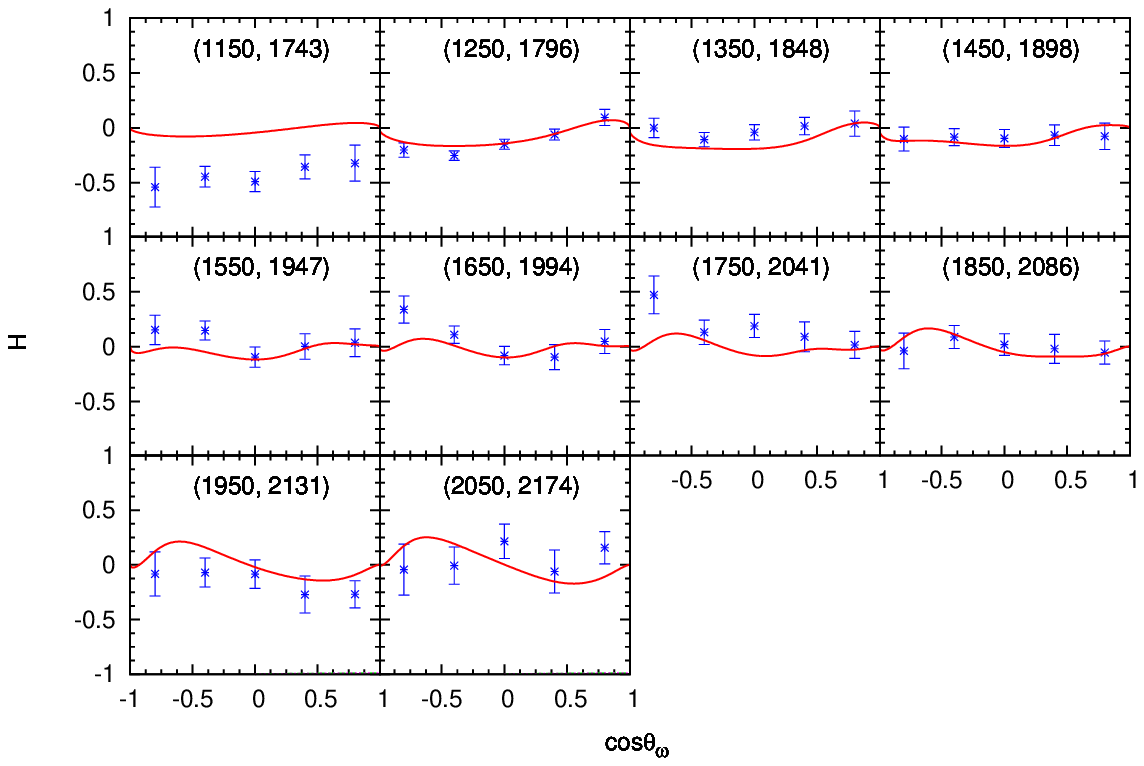}
\caption{Beam-target asymmetry $H$ for $\gamma p \rightarrow \omega p$ as a function of $\cos\theta_\omega$ in the center-of-mass frame. Numbers in parentheses are defined in Fig.~\ref{pic:dsig} caption. Data are taken from CLAS Collaboration \cite{Roy:2018}. }
\label{pic:h}
\end{figure*}

\begin{figure}[htbp]
\includegraphics[width=0.85\columnwidth]{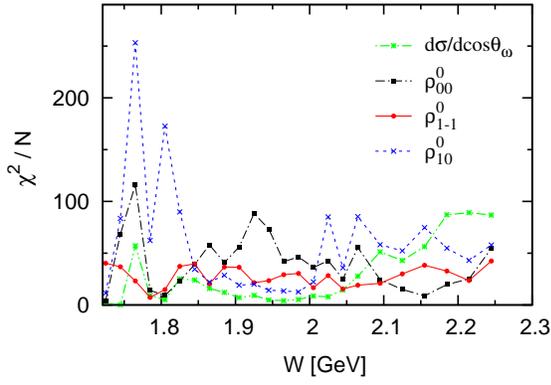}
\caption{$\chi^2/N_i$ versus the center-of-mass energy of the system for $i=$ $d\sigma$, $\rho^0_{00}$, $\rho^0_{1-1}$ and Re\,$\rho^0_{10}$.  }
\label{pic:histogram}
\end{figure}

\begin{table*}[htbp]
\caption{\label{table:chi2} $\chi_i^2/N_i$ evaluated for a given type of observable specified by the index $i=d\sigma$ (differential cross section), $\rho^0_{00}$, $\rho^0_{1-1}$, Re\,$\rho^0_{10}$, $\Sigma$, $T$, $P'$, $E$, $F$ and $H$. The last column corresponds to the global $\chi^2 / N$, where $N$ is the total number of data points including all the types of observables considered. Row ``sol. I" corresponds to the best fit results with the parameters presented in Tables~\ref{table:decay channel}$-$\ref{table:mass and width}. Row ``sol. II" corresponds to the fit results with $\Lambda_t=650$ MeV. Row ``sol. III" corresponds to the same fit results as ``sol. I", except for the presence of the fit phase parameter $\varphi$ $(=0.05\pi)$ in the complex exponential factor $e^{{\rm i}\varphi}$ in the non-resonant amplitude.}
\renewcommand{\arraystretch}{1.2}
\begin{tabular*}{\textwidth}{@{\extracolsep\fill}lccccccccccc}
\hline\hline
       & $\chi_{d\sigma}^2/N_{d\sigma}$ & $\chi_{\rho_{00}}^2/N_{\rho_{00}}$ & $\chi_{\rho_{1-1}}^2/N_{\rho_{1-1}}$ & $\chi_{\rho_{10}}^2/N_{\rho_{10}}$ & $\chi_{\Sigma}^2/N_\Sigma$ & $\chi_T^2/N_T$ & $\chi_{P'}^2/N_{P'}$ & $\chi_{E}^2/N_E$ &  $\chi_{F}^2/N_{F}$ & $\chi_{H}^2/N_{H}$ & $\chi^2/N$  \\
       &$N_{d\sigma}=402$ &$N_{\rho_{00}}=402$ &$N_{\rho_{1-1}}=402$ &$N_{\rho_{10}}=402$ &$N_\Sigma=81$ &$N_T=95$ &$N_{P'}=50$ &$N_{E}=88$ &$N_{F}=99$ &$N_{H}=50$ &$N=2071$ \\   \hline
sol. I   &  $28.7$  & $40.6$  & $27.6$  &  $57.7$  & $2.8$  & $5.1$  & $2.6$ & $1.9$  &  $3.0$ &  $2.5$ & $30.7$ \\
sol. II  &  $31.1$  & $46.0$  & $23.4$  &  $56.0$  & $3.5$  & $5.1$  &  $2.5$ & $2.4$  &  $2.9$ &  $2.6$ & $31.1$  \\
sol. III  &  $28.2$ & $38.1$  & $28.4$  &  $58.4$  & $2.6$  & $4.6$  &  $2.7$ &  $1.8$ &  $3.1$ &  $2.7$  & $30.4$  \\
\hline\hline
\end{tabular*}
\end{table*}

In Table~\ref{table:decay channel} we list the decay channels and the corresponding branching ratios we have considered for each of the resonances. They are fixed to be the centroid values of the dominant decay modes quoted in PDG \cite{PDG:2018}, with few exceptions. 
For the $N(1520)3/2^-$ resonance, the centroid value of the $N\pi\pi$ branching ratio quoted in PDG is $\beta_{N\pi\pi} = 0.30$. However, we adopt the value of $\beta_{N\pi\pi}\cong 0.12$ to satisfy the sum rule given by Eq.~(\ref{eq:sumrule}), as we consider the $N\pi\pi$ decay mode as an effective mode to account for those not included in the present model.  Likewise, for the other resonances, where the sum of the centroid values of the corresponding branching ratios quoted in PDG exceeds the sum rule, these centroid values, except for the $N\pi$ branching ratio, are reduced by a common factor (for a given resonance) which is determined by the fit to the data to fulfill the sum rule. We have checked that the calculated results are not much sensitive to the particular values used for the branching ratios in question of these resonance decay modes. Since no information on the $N\gamma$ branching ratio is given in PDG for the $N(1860)5/2^+$ resonance, here, it is simply taken to be $\beta_{N\gamma}=0.05$.

For all the above-threshold resonances, the $N\omega$ branching ratios, $\beta_{N\omega}$, are calculated from the corresponding fitted $RN\omega$ coupling constants, $g^{(1,2,3)}_{RN\omega}$. They are also displayed in Table~\ref{table:decay channel}. For comparison, the range of the corresponding values given in PDG \cite{PDG:2018} are also shown in square brackets. We see that, overall, the extracted values of $\beta_{N\omega}$ are consistent with those quoted in PDG.  Our value for the $N(1720)3/2^+$ resonance lay far outside the corresponding range given in PDG.  We can easily bring it within the range given in PDG by tweaking the corresponding $N\gamma$ branching ratio within its range as displayed in Table~\ref{table:decay channel} with no significant change in the results for the observables considered. Indeed,  if we take $\beta_{N\gamma}=0.05$, we obtain $\beta_{N\omega}=12.98 \pm 0.02$. In this case, the fitted $N(1720)3/2^+$ mass and width turn out to be $M_R = 1750 \pm 2$ MeV and $\Gamma_R = 140 \pm 3$, respectively.

\begin{figure*}[htbp]
\includegraphics[width=0.75\textwidth]{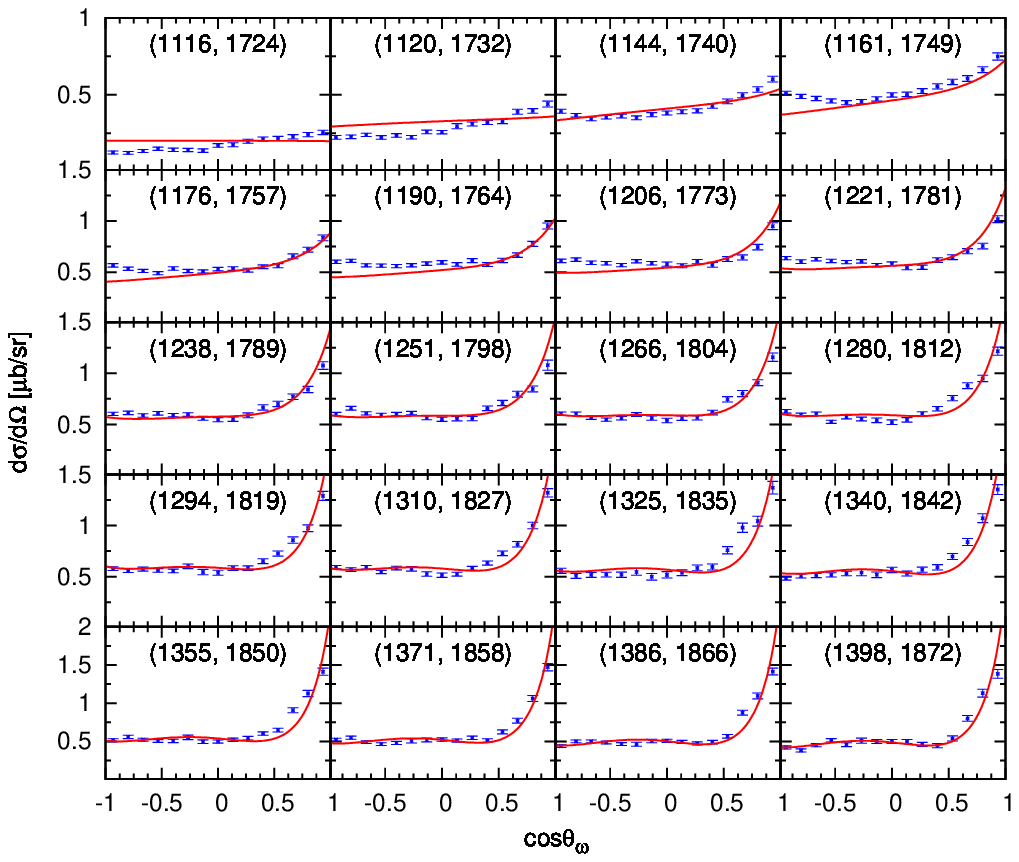}
\caption{Differential cross section for $\gamma p \rightarrow \omega p$ as a function of $d\cos\theta_\omega$ compared to the data from the A2 Collaboration \cite{Strakovsky:2014wja}. Numbers in parentheses are defined in Fig.~\ref{pic:dsig} caption. These data are not included in the fit.}
\label{pic:dsigmami}
\end{figure*}

\begin{figure*}[htbp]
\includegraphics[width=0.75\textwidth]{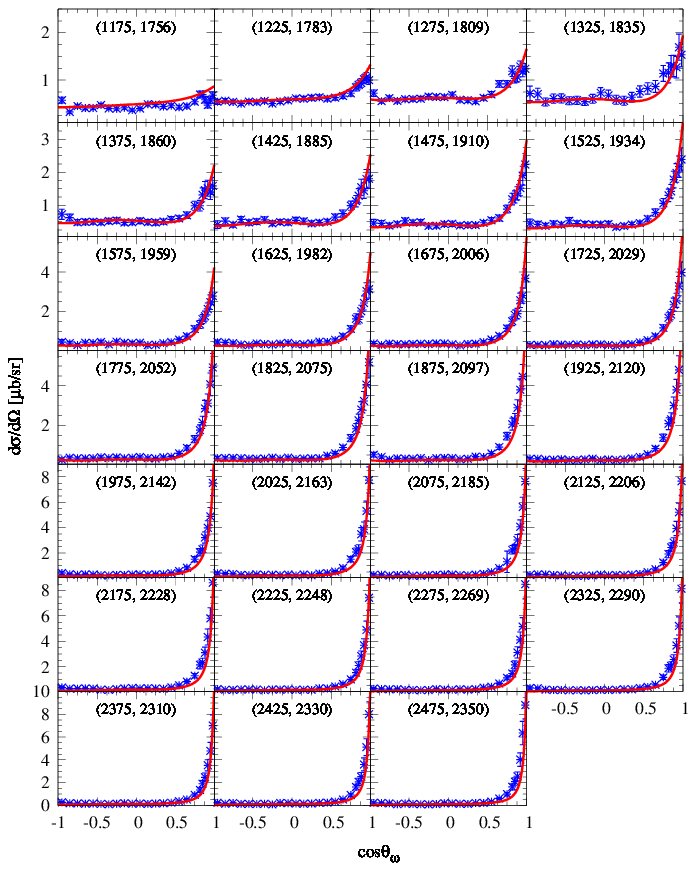}
\caption{Differential cross section for $\gamma p \rightarrow \omega p$ as a function of $d\cos\theta_\omega$  compared to the data from the CBELSA/TAPS Collaboration \cite{Denisenko:2016ugz}. Numbers in parentheses are defined in Fig.~\ref{pic:dsig} caption. These data are not included in the fit.}
\label{pic:dsigcbel}
\end{figure*}

\begin{figure*}[htbp]
\includegraphics[width=0.75\textwidth]{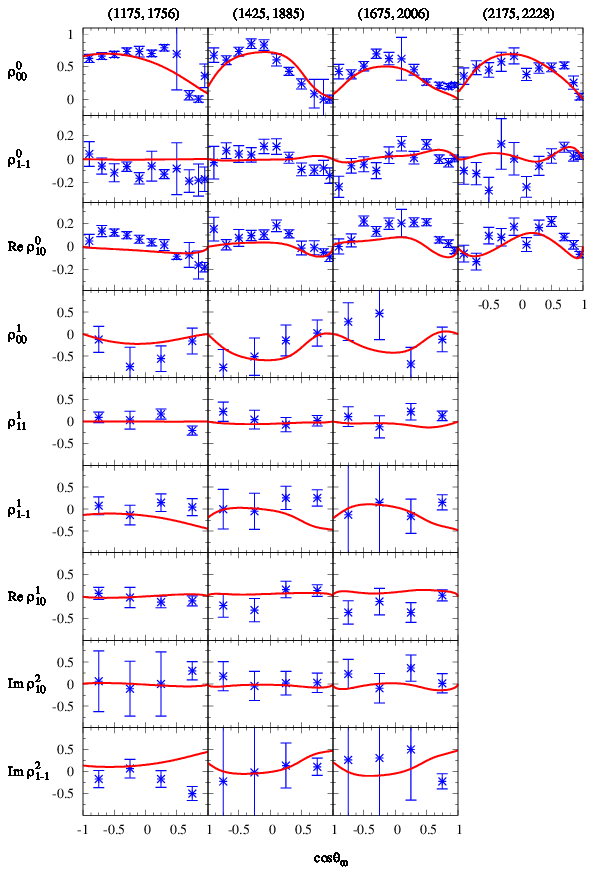}
\caption{Spin Density Matrix Elements in the Adair frame for $\gamma p \rightarrow \omega p$ as a function of $d\cos\theta_\omega$ compared to the data from the CBELSA/TAPS Collaboration \cite{Denisenko:2016ugz}. Numbers in parentheses are defined in Fig.~\ref{pic:dsig} caption. These data are not included in the fit. }
\label{pic:rhocbel}
\end{figure*}

Given the limitations involved in the extraction of the branching ratios in tree-level type calculations as mentioned earlier, the present results are quite reasonable estimations of the $N\omega$ branching ratios. As has been pointed out, only the product of the vertex functions are well-defined in the present-type calculations. Hence, one should take our values for the $\beta_{N\omega}$ with a considerable grain of salt.

In Table~\ref{table:constants}, we show the resulting fit values of the product of the coupling constants, $g^{(i)}_{RN\gamma} g^{(j)}_{RN\omega}$, and the phase $\phi_R$ common to the displayed products of the coupling constants for each resonance as explained previously. As mentioned earlier, note that in a tree-level Lagrangian approach, only the product of the vertex functions  are well constrained.

The $N(1520)3/2^-$, $N(1700)3/2^-$ and $N(1720)3/2^+$ resonances have been also considered in Ref.~\cite{Titov:2002iv} within an effective Lagrangian approach in which the resonance couplings are fixed from the empirical helicity amplitudes together with the vector meson dominance assumption. Note that only the terms with $g^{(1)}_{RN\gamma}*g^{(1)}_{RN\omega}$ in the resonance amplitudes have been considered in Ref.~\cite{Titov:2002iv} and these products turn out to be smaller than those obtained in the present work. As a consequence, the $N(1520)3/2^-$ resonance, in particular, is not prominent in Ref.~\cite{Titov:2002iv} in contrast to the finding of this work. For the $N(1700)3/2^-$ and $N(1720)3/2^+$ resonances, their couplings in Ref.~\cite{Titov:2002iv} are fixed from the empirical helicity amplitudes quoted in the PDG of 2001 and most of these values are about an order of magnitude smaller than those in the PDG of 2019 \cite{PDG:2018}.

In Table~\ref{table:cutoff}, we show the resulting fit values for the cutoff parameters $\Lambda_t$, $\Lambda_s$ and $\Lambda_u$ of the form factors in the $t$-, $s$-, and $u$-channel non-resonant amplitudes. Here we note that the value of $\Lambda_u=500$ MeV entering in the nucleon $u$-channel diagram may be too small to be realistic, indicating that the nucleon current contribution is too large and, therefore, a strong form factor is needed to suppress its contribution. This feature has been observed also in the earlier works \cite{Oh:2000zi,Oh:2002,Paris:2009}. In Ref.~\cite{Paris:2009}, where the analysis has been carried out within a coupled channel approach, instead of suppressing the too strong nucleonic current contribution through a form factor, it has introduced a fictitious heavier ``omega" meson, $\omega'$, with an adjustable coupling constant. The $NN\omega$ coupling constants, $g_{NN\omega}$ and $\kappa_\omega$ are not yet well determined. An early account in Ref.~\cite{Nakayama:1998} reveals a broad range of values from $g^2_{NN\omega}/4\pi \sim 8$ to 35 for the vector coupling, and $\kappa_\omega \sim -0.16$ to +0.14 for the ratio of the tensor to vector coupling constants. Janssen \textit{et al.} \cite{Janssen:1996} have shown that, once the contribution of the correlated $\pi\rho$-exchange to the $NN$ interaction is taken into account explicitly, the large values of $g^2_{NN\omega}/4\pi \cong 20$ required in the description of the $NN$ scattering data is reduced by about a factor of $2$, leading to an $NN\omega$ coupling constant which is more in line with the value one would obtain from the SU(3) symmetry considerations, $g_{NN\omega} = 3 g_{NN\rho}  \approx 10$. This reduces the range of $g_{NN\omega}$ considerably. The value of $g_{NN\omega} \approx 10$, consistent with SU(3), have been  used in the analyses of the $NN \to NN \omega$ reaction \cite{Nakayama:1998,Nakayama:2007}. Also, the values of $g_{NN\omega}=7.0-10.5$ and $\kappa_\omega\approx 0$ were found to describe consistently the $\pi N$ scattering and $\pi$ photoproduction processes \cite{Sato:1996gk}. Anyway, the problem of a too strong nucleonic current, especially for the $u$-channel contribution, is an open question and should be addressed in future work. Here, together with the $t$-channel pion exchange current, the nucleonic current has been introduced merely as to account for the non-resonant background amplitude, since the focus of this work is to extract information on the resonance content in $\omega$ photoproduction reaction.

The fit results for the resonance mass $M_R$, total width $\Gamma_R$ at $W=M_R$ and the cutoff parameter $\Lambda_R$ are given in Table~\ref{table:mass and width}, together with the corresponding range of values quoted in PDG \cite{PDG:2018} (in square brackets). Overall, our extracted values of both $M_R$ and $\Gamma_R$ are consistent with those quoted in PDG \cite{PDG:2018}. The extracted masses of the $N(1875)3/2^+$ and $N(1895)1/2^-$ resonances as well as the total width of the latter are somewhat larger than those quoted in PDG.

\begin{figure}[htbp]
\includegraphics[width=0.85\columnwidth]{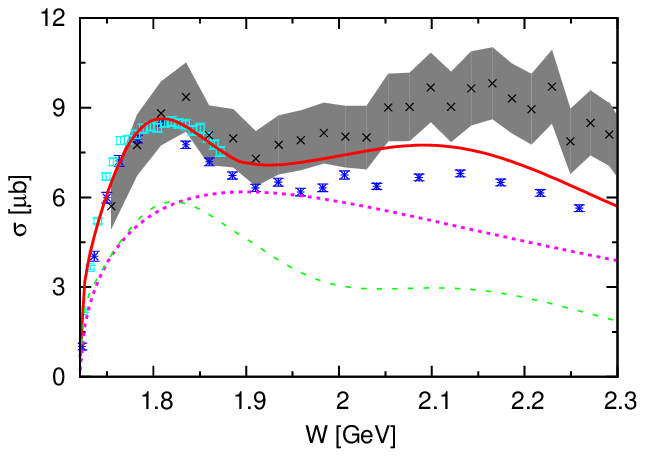}
\caption{Total cross section for $\gamma p \rightarrow \omega p$ as a function of center mass energy for the whole system. The solid, dotted and dashed lines denote the contributions from the full amplitude, the $t$-channel $\pi$ exchange and the $s$-channel resonance exchanges, respectively. The contributions from other terms are too small to be shown in this figure. The scattered symbols denote the data from the SAPHIR Collaboration \cite{Barth:2003kv} (star), the A2 Collaboration \cite{Strakovsky:2014wja} (open square) and the CBELSA/TAPS Collaboration \cite{Wilson:2015uoa} (cross), respectively, with the grey band representing the systematic uncertainties. These data are not included in the fit.}
\label{pic:total1}
\end{figure}

\begin{figure}[htbp]
\includegraphics[width=0.85\columnwidth]{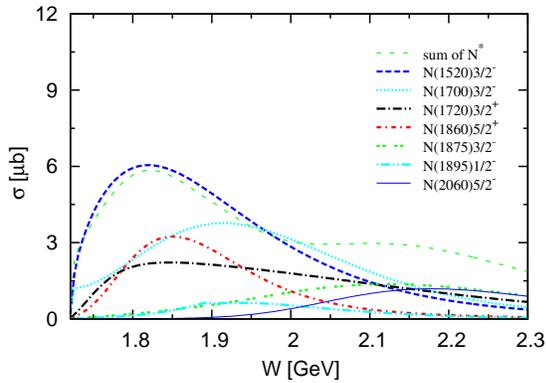}
\caption{Total cross section for $\gamma p \rightarrow \omega p$ from the contributions of individual resonances as a function of center mass energy for the whole system.}
\label{pic:total2}
\end{figure}

\begin{figure}[htbp]
\includegraphics[width=0.85\columnwidth]{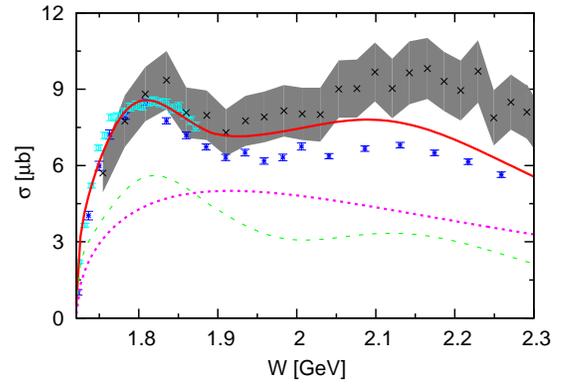} \\[6pt]
\includegraphics[width=0.85\columnwidth]{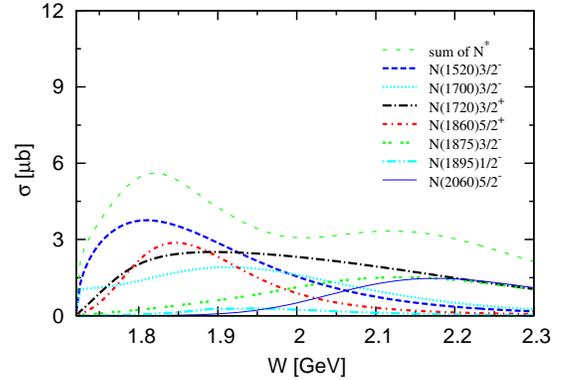}
\caption{Total cross section results for $\gamma p \rightarrow \omega p$ with a reduced background contribution by choosing the $t$-channel cutoff parameter $\Lambda_t=650$ MeV instead of $700$ MeV which is obtained from a fit to the CLAS data. The upper panel shows the contributions from the full amplitude, the $t$-channel $\pi$ exchange and the $s$-channel resonance exchanges. The lower panel shows the contributions from the individual resonances. The notations are the same as those of Fig.~\ref{pic:total1} for the upper panel and as those of Fig.~\ref{pic:total2} for the lower panel.}
\label{pic:total12}
\end{figure}

\begin{figure}[htbp]
\includegraphics[width=0.85\columnwidth]{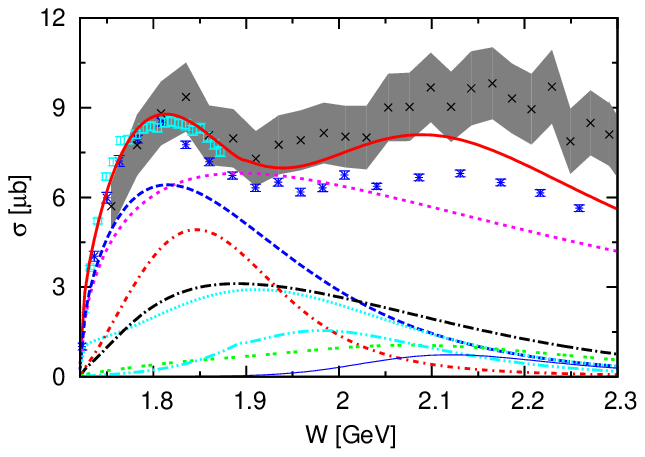} \\[6pt]
\includegraphics[width=0.85\columnwidth]{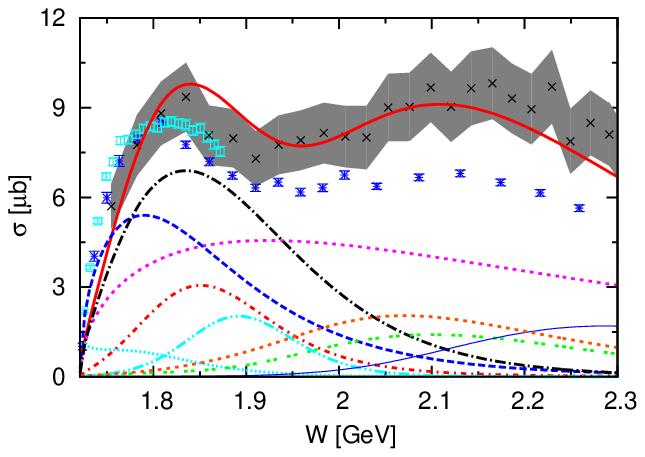}
\caption{Total cross section results from the fits to the A2 data (upper panel) and the CBELSA/TAPS data (lower panel) as explained in the text. For notations of the panels, see Figs.~\ref{pic:total1} and \ref{pic:total2}. Note that the fit results of the CBELSA data contains an extra resonance, $N(2120)3/2^-$ (the orange dotted line) compared to the fit results of the CLAS and A2 data.}
\label{pic:total123_A2_CBELSA}
\end{figure}

In Figs.~\ref{pic:dsig}-\ref{pic:h} we show our numerical results for differential cross section $d\sigma/d\cos\theta$, spin density matrix elements $\rho^0_{00}$, ${\rm Re}\,\rho^0_{10}$ and $\rho^0_{1-1}$, single polarization $\Sigma$ and $T$, and beam-target double polarization $P'$, $E$, $F$ and $H$, respectively, together with the corresponding most recent experimental data.  Overall, the agreement of the model results with the data is reasonable. In the differential cross sections, deviations in the fit results are seen at the backward angles and low energies.
One also sees significant deviations in the description of the SDMEs by the present model, especially, ${\rm Re}\,\rho^0_{10}$ for forward angles at higher energies and $\rho^0_{00}$. Some deviations are also observed in $\rho^0_{1-1}$ at forward and backward angles depending on the energy region. Except for $E$ and $F$, there are some deviations for other spin observations at the lowest energies. Also, there is a discrepancy in $E$ for forward angles at $W=1898$ MeV.

The chi squared per data points for different types of observables are shown in Table~\ref{table:chi2} (row ``sol. I").  We see that the relatively large global $\chi^2/N$ obtained arises, especially, from the SDMEs $\rho^0_{10}, \rho^0_{00}$ and $\rho^0_{1-1}$, followed by the differential cross section. This is caused by the very small statistical uncertainties in these data. Note that, here, no systematic uncertainties are included.

When the systematic uncertainties are taken into account in calculating $\chi^2$, the $\chi^2/N_i$ values for $d\sigma$, $\rho^0_{00}$, $\rho^0_{1-1}$ and Re\,$\rho^0_{10}$ are $1.4$, $10.9$, $16.6$, and $21.6$, respectively. The $\chi^2/N_i$ values for $\rho^0_{00}$, $\rho^0_{1-1}$ and Re\,$\rho^0_{10}$ are still relatively large as both the  statistical and systematic uncertainties in these data are small.  To illustrate the energy dependence of $\chi^2/N_i$, the $\chi^2/N_i$ versus the center-of-mass energy of the system $W$ for $i=$ $d\sigma$, $\rho^0_{00}$, $\rho^0_{1-1}$ and Re\,$\rho^0_{10}$ are shown in Fig.~\ref{pic:histogram}. As can be seen, the largest contribution to $\chi_{\rho_{10}}^2/N_{\rho_{10}}$ comes, by far, from that corresponding to  $W=1765$ MeV, followed by those at $W=1805$ MeV and $W > 2000$ MeV.  To $\chi_{\rho_{00}}^2/N_{\rho_{00}}$, the larger contributions arise from those corresponding to energies $W = 1765$ MeV and  $[1860 \sim 2065]$ MeV, whereas, to $\chi_{\rho_{1-1}}^2/N_{\rho_{1-1}}$, the contributions are rather constant as a function of energy. The relatively large contribution to $\chi^2/N_{d\sigma}$ arise not only from that corresponding to $W=1765$ MeV but also from those at higher energies of $W \ge 2100$ GeV. Note that $\chi^2/N_{d\sigma}$ increases with energy for $W > 2000$GeV, indicating that the present model may lack some reaction mechanism(s) at these higher energies.  A potential candidate may be the Pomeron exchange.  At the lowest energy, the agreement with the cross section data can be improved if the finite width of the $\omega$ meson is taken into account. 

In Figs.~\ref{pic:dsigmami} and \ref{pic:dsigcbel}, the theoretical differential cross sections resulted from the fit to the CLAS data are compared with the data from the A2 Collaboration \cite{Strakovsky:2014wja} and the data from the CBELSA/TAPS Collaboration \cite{Denisenko:2016ugz}, respectively. One sees significant deviations between the theoretical results and the A2 data, especially, for backward angles at lower energies. One also sees obvious deviations between the theoretical results and the CBELSA/TAPS data in the forward angle region at higher energies. These deviations clearly indicate the discrepancies among the data from the CLAS, A2 and CBELSA/TAPS Collaborations as mentioned at the beginning of this section. In Fig.~\ref{pic:rhocbel}, the theoretical results for the SDMEs resulted from a fit to the CLAS data are compared with the corresponding data from the CBELSA/TAPS Collaboration \cite{Denisenko:2016ugz}. Here, no obvious discrepancies are seen, mainly due to the fact that the data from the CBELSA/TAPS Collaboration has much larger error bars.

The prediction of the total cross section from the present model is shown in Fig.~\ref{pic:total1}. There, the solid curve corresponds to theoretical total cross sections obtained by integrating the corresponding differential cross sections as shown in Fig.~\ref{pic:dsig}. The dotted and dashed curves represent the contributions from the $t$-channel $\pi$ exchange and the $s$-channel resonance exchanges, respectively. We see that the contribution from the $\pi$ exchange is significant in the whole energy region considered and it becomes dominant for energies above $W \sim 1.9$ GeV. The contributions from the resonances are also significant. In Fig.~\ref{pic:total1} we also display the available total cross section data. The stars and open squares correspond to the data from the SAPHIR Collaboration \cite{Barth:2003kv} and the A2 Collaboration \cite{Strakovsky:2014wja}, respectively. The crosses represent the CBELSA/TAPS data.  Here we see clearly the currently existing consistencies/inconsistencies among these data. Due to the limited angular acceptance of the CLAS detector, no total cross section data exist from this Collaboration. However, the solid curve gives some idea on where the corresponding data might lay.

In Fig.~\ref{pic:total2}, we show the total cross sections stemming from the contributions of individual resonances. One sees that the energy region close to threshold is dominated by the $N(1520)3/2^-$ resonance which causes a sharp rise of the cross section from the threshold and peaks around $W \sim 1.8$ GeV. The next strongest contribution is due to the $N(1700)3/2^-$ resonance, followed by the $N(1860)5/2^+$ and $N(1720)3/2^+$  resonances. Both the $N(1700)3/2^-$ and $N(1720)3/2^+$ exhibit a rather broad contribution, while the $N(1860)5/2^+$  resonance contribution peaks around $W \sim 1.85$ GeV. Note that the nominal $N\omega$ threshold is $W_{\rm thres} \cong 1721$ MeV. The higher mass resonances, $N(1875)3/2^-$ and $N(1895)1/2^-$, become relevant for energies above $W \sim 1.9$ GeV, and the $N(2060)5/2^-$ resonance, above $W \sim 2$ GeV.

In order to gain some insight on how well the considered CLAS data constrain the background and resonance contributions, we make a comparison of the present model results shown in Figs.~\ref{pic:dsig}-\ref{pic:h} and \ref{pic:total1}-\ref{pic:total2} with other fit results that have a little worse but comparable fit quality and have a reduced  background contribution. This is achieved by reducing the $\pi$ exchange contribution -- by far the dominant non-resonant background contribution -- through the reduced cutoff parameter $\Lambda_t$ (the only free parameter in the $\pi$ exchange contribution) from $\Lambda_t = 700$ MeV to $\Lambda_t=650$ MeV. The corresponding total cross section results are shown in Fig.~\ref{pic:total12}. Comparing with the results from the best fit with $\Lambda_t = 700$ MeV as shown in Figs.~\ref{pic:total1}-\ref{pic:total2}, we see that the reduction in the background contribution can be compensated by the enhancement in the resonance contribution and still achieve a reasonable fit quality of the considered data (cf. the upper panel in Fig.~\ref{pic:total12}). We also see that, although the resonance content remain the same, the relative contribution of the individual resonances changes (cf. the lower panel in Fig.~\ref{pic:total12}).

The chi squared per data points for different types of observables corresponding to the fit results with $\Lambda_t=650$ MeV are shown in Table~\ref{table:chi2} (row ``sol. II") which also give an idea of the comparable fit quality to that of the best fit results with $\Lambda_t=700$ MeV corresponding to ``sol I" in Table~\ref{table:chi2}.

In a full reaction amplitude, both the resonant  and non-resonant background amplitudes are, in general, complex. The complex nature of the amplitude is crucial for describing certain spin observables in particular. In the present model, the non-resonant amplitude is purely real. A rough estimate of the relevance/irrelevance of  the complex nature of the non-resonant amplitude may be obtained by introducing a complex phase factor $e^{{\rm i}\phi_{NR}}$ in our real non-resonant amplitude. Of course, the complex  structure of the reaction amplitude is intimately related to the property of unitarity of the full reaction amplitude. In other words,  unitary should, in principle, dictate the complex phases.\footnote{See Ref.~\cite{Razavi:2019}, where the complex phase structures of the full meson-baryon and photoproduction amplitudes have been exposed.} We have then repeated the fit of the CLAS data as has been done to obtain the fit results corresponding to ``sol. I" in Table~\ref{table:chi2} with the phase $\phi_{NR}$ as an extra fit parameter. The results obtained for $\chi^2_i/N_i$ for the individual independent observables $i$ are shown in Table~\ref{table:chi2} (row ``sol. III"). Comparing with the results of row ``sol. I", we see that this can influence the fit quality of certain spin observables by as much as $\sim 10\%$. 

We now turn our attention to the analysis of the A2 \cite{Strakovsky:2014wja} and CBELSA/TAPS  \cite{Wilson:2015uoa} data to see how they influence the extracted resonance content compare to the CLAS data. Recall that there are some inconsistencies in the cross section of both the A2 and CBELSA/TAPS data with the corresponding CLAS data as discussed at the beginning of this section.  For the analysis of the A2 data, we have simply replaced the CLAS differential cross section data by the corresponding A2 data in the overlapping energy region. Everything else are kept as in the analysis of the CLAS data presented above. For the analysis of the CBELSA/TAPS data, we consider the differential cross section and SDMEs $\rho^0_{00}$, ${\rm Re}\,\rho^0_{10}$, $\rho^0_{1-1}$ \cite{Wilson:2015uoa}, the spin observables $\Sigma$ \cite{Klein:2008aa}, $E$ and $G$ \cite{Eberhardt:2015lwa}. For the A2 Collaboration data shown in Fig.~\ref{pic:dsigmami} with 300 data points and CBELSA/TAPS Collaboration data shown in Fig.~\ref{pic:dsigcbel} with 648 data points, the $\chi^2_{d\sigma}/N_{d\sigma}$ are $3.3$ and $2.6$, respectively. The total cross section results from the fits to the A2 and the CBELSA/TAPS data as explained above are shown in Fig.~\ref{pic:total123_A2_CBELSA}. A comparison of Figs.~\ref{pic:total1}-\ref{pic:total2} with the upper panel of Fig.~\ref{pic:total123_A2_CBELSA} reveals how the use of the A2 differential cross section data instead of the corresponding CLAS data in the resonance extraction affects the results. We see that the resonances required to reproduce the data from both groups remain unchanged, but the relative contribution of these resonances changes, in particular, of the resonances $N(1520)3/2^-$ and $N(1860)5/2^+$. Comparing Figs.~\ref{pic:total1}-\ref{pic:total2} with the lower panel of Fig.~\ref{pic:total123_A2_CBELSA}, we also see a different relative contribution of the resonances between the CLAS and CBELSA/TAPS data. Note in particular that at least an extra higher mass resonance -- here, $N(2120)3/2^-$ -- is required to reproduced the CBELSA/TAPS differential cross section data for energies $W > 2$ GeV in contrast to the analysis of the CLAS data.

\section {Summary and conclusion} {\label{summary}}

Quite recently, the CLAS Collaboration has reported the newest high-precision data on the spin observables $E$ \cite{Akbar:2017uuk} as well as $\Sigma$, $T$ \cite{Roy:2017qwv}, and $P'$, $F$, $H$ \cite{Roy:2018} for the $\gamma p \to \omega p$ photoproduction reaction. The data for the latter four observables, $T$, $P'$, $F$ and $H$, were measured for the very first time.

In this work, we have performed for the $\gamma p \to \omega p$ reaction a theoretical analysis of the recently published high-precision data on spin observables $\Sigma$, $T$, $P'$, $E$, $F$ and $H$ \cite{Akbar:2017uuk,Roy:2017qwv,Roy:2018} together with the previously published high-precision data on differential cross sections and SDMEs $\rho^0_{00}$, $\rho^0_{1-1}$, Re\,$\rho^0_{10}$ \cite{Williams:2009ab} from the CLAS Collaboration within an effective Lagrangian approach. Part of the results for the double polarization observables $P'$, $F$ and $H$ have been published together with the experimental data in Ref.~\cite{Roy:2018}. Here we have reported the details of our investigations, showing, in particular, the results for $\Sigma$, $T$, $E$, $d\sigma/d\Omega$ and SDMEs.

In the present work, the reaction amplitudes were constructed by considering the $t$-channel $\pi$ and $\eta$ exchanges, the $s$-channel nucleon and nucleon resonances exchanges, the $u$-channel nucleon exchange and the generalized contact current. The generalized contact current is formulated in such a way that the full photoproduction amplitudes satisfy the generalized Ward-Takahashi identity and thus are fully gauge invariant. The $s$-channel nucleon resonances were introduced as few as possible to get a satisfactory description of the data.

It has been shown that all the available data from the CLAS Collaboration can be satisfactorily described by considering the $N(1520)3/2^-$, $N(1700)3/2^-$, $N(1720)3/2^+$, $N(1860)5/2^+$, $N(1875)3/2^-$, $N(1895)1/2^-$ and $N(2060)5/2^-$ resonances in the $s$-channel apart from the non-resonant contributions. The masses and widths for these resonances have been extracted and compared with those quoted by PDG \cite{PDG:2018}. They are in good agreement overall. Although a proper extraction of the branching ratios requires a coupled channels approach, we have also estimated these quantities within the present model. The contributions from the individual terms of the reaction amplitudes to the total cross sections have been analyzed. The $t$-channel $\pi$ exchange is found to dominate the background contribution, and the contributions from the nucleon resonances were also found to be significant to the cross sections. The effects of the data from the A2, GRALL, SAPHIR and CBELSA/TAPS Collaborations to the resonance content extracted from the CLAS data for this reaction have been discussed.

\begin{acknowledgments}
The authors N.~C.~Wei and F.~Huang are grateful to Dr. Q.~F.~L{\"u} for his useful discussions during the early stage of this work. The authors also thank Prof. V.~Crede for providing us the experimental data. This work is partially supported by the National Natural Science Foundation of China under Grant No.~11475181 and No.~11635009, the Youth Innovation Promotion Association of CAS under Grant No.~2015358, and the Key Research Program of Frontier Sciences of CAS under Grant No. Y7292610K1.
\end{acknowledgments}

\end{document}